\documentclass[12pt]{article}
\usepackage{amsmath,amsthm,amssymb,epsfig,euscript,array,cancel,color}
\usepackage[colorlinks=true]{hyperref}
\usepackage{verbatim}
\usepackage{url}
\usepackage{bbm}

\usepackage[paper=a4paper,dvips,top=2.54cm,left=2.74cm,right=2.34cm,
    foot=1cm,bottom=2.54cm]{geometry}

\newcounter{multieqs}

\newcommand{\be}{\begin{equation}}
\newcommand{\ee}{\end{equation}}
\newcommand{\eq}[1]{(\ref{#1})}
\newcommand{\bit}{\begin{itemize}}  \newcommand{\eit}{\end{itemize}}

\def\bd{\begin{document}}
\def\ed{\end{document}}
\def\nn{\nonumber}
\def\bea{\begin{eqnarray}}
\def\eea{\end{eqnarray}}
\let\bm=\bibitem

\def\a{\alpha}      \def\da{{\dot\alpha}}  \def\dA{{\dot A}}
\def\b{\beta}       \def\db{{\dot\beta}}  
\def\g{\gamma}  \def\G{\Gamma}  \def\dc{{\dot\gamma}}  
\def\d{\delta}  \def\D{\Delta}  \def\ddt{\dot\delta}  
\def\e{\epsilon}        \def\ve{\varepsilon}  
\def\f{\phi}    \def\F{\Phi}    \def\vvf{\f}  
\def\h{\eta}  
\def\k{\kappa}  
\def\l{{\lambda}} \def\L{\Lambda}  
\def\m{\mu} \def\n{\nu}  
\def\o{\omega}  
\def\p{\pi} \def\P{\Pi}  
\def\r{\rho}  
\def\s{\sigma}  \def\S{\Sigma}  
\def\t{\tau}  
\def\th{\theta} \def\Th{\Theta} \def\vth{\vartheta}  
\def\X{\Xeta}  
\def\z{\zeta}  

\def\na{\nabla}

\def\cA{{\cal A}} \def\cB{{\cal B}} \def\cC{{\cal C}}  
\def\cD{{\cal D}} \def\cE{{\cal E}} \def\cF{{\cal F}}  
\def\cG{{\cal G}} \def\cH{{\cal H}} \def\cI{{\cal I}}  
\def\cJ{{\cal J}} \def\cK{{\cal K}} \def\cL{{\cal L}}  
\def\cM{{\cal M}} \def\cN{{\cal N}} \def\cO{{\cal O}}  
\def\cP{{\cal P}} \def\cQ{{\cal Q}} \def\cR{{\cal R}}  
\def\cS{{\cal S}} \def\cT{{\cal T}} \def\cU{{\cal U}}  
\def\cV{{\cal V}} \def\cW{{\cal W}} \def\cX{{\cal X}}  
\def\cY{{\cal Y}} \def\cZ{{\cal Z}}

\def\ua{{\underline{\alpha}}} 
\def\ub{{\underline{\phantom{\alpha}}\!\!\!\beta}}
\def\uc{\underline{\phantom{\alpha}}\!\!\!\gamma}  
\def\um {{\underline{\mu}}}  
\def\ud{{\underline{\delta}}} 
\def\ue{\underline\epsilon}  
\def\una{{\underline a}}\def\uA{{\underline A}}  
\def\unb{{\underline b}}\def\uB{{\underline B}} 
\def\unc{{\underline c}}\def\uC{{\underline C}}  
\def\und{{\underline d}}\def\uD{{\underline D}}  
\def\une{{\underline e}}\def\uE{{\underline E}}  
\def\unf{{\underline{\phantom{e}}\!\!\!\! f}}\def\uF{{\underline F}}  
\def\unm{{\underline m}\def\uM{\underline M}} 
\def\unn{{\underline n}\def\uN{\underline N}} 
\def\unp{{\underline{\phantom{a}}\!\!\! p}}\def\uP{{\underline P}}  
\def\unq{{\underline{\phantom{a}}\!\!\! q}}  
\def\uQ{{\underline{\phantom{A}}\!\!\!\! Q}}  
\def\uH{{\underline{H}}}  
\def\uM{{\underline{M}}}
\def\uN{{\underline{N}}}
\def\unl{{\underline{l}}}

\def\Ft{\tilde{F}}
\def\Gt{\tilde{G}}
\def\Ht{\tilde{H}}

\newcommand{\ov}[1]{\overrightarrow{#1}}

\def\pa{\partial} \def\del{\partial}  
  
\newcommand{\tr}{\mbox{tr}}

\def\hlf{\frac{1}{2}}  
\def\ove#1{\frac{1}{#1}}

\def\RR{\mathbb{R}}

\newcommand{\ex}[1]{{\rm e}^{#1}} \def\ii{{\rm i}}

\newcommand{\lrbrk}[1]{\left(#1\right)}

\def \diag {\text{diag}}

\def\w{{\wedge}}

\numberwithin{equation}{section}

\author{Pichet Vanichchapongjaroen\footnote{pichetv@nu.ac.th}$~^a$
\\
\\
{\small $^a$ \it The Institute for Fundamental Study ``The Tah Poe Academia Institute",}
\\
{\small\it Naresuan University, Phitsanulok 65000, Thailand}}

\title{Duality-symmetric D3-brane action with twisted self-dual 2-form doublet}

\begin{document}
\maketitle

\abstract{In this paper we apply the Sen formalism, which is originally developed for chiral form fields,
to construct non-linear $SL(2,\mathbb{R})$ duality-symmetric actions in four dimensions.
The non-linear actions contain a potential term
whose allowed form satisfies a condition arising from a requirement that twisted self-duality condition of the theory is equivalent to a constitutive relation. This condition can be perturbatively solved to obtain the potential term.
In special cases such as DBI theory or D3-brane, the potential term has a closed form.
In particular, we construct the $SL(2,\mathbb{R})$ duality-symmetric D3-brane action coupled to the type IIB supergravity background in the Sen formalism.
Key features of the Sen formalism are also present in the duality-symmetric actions.
For example, there are unphysical fields with the wrong sign of kinetic terms.
These fields are uncoupled from physical fields at the equations of motion level.
We show that the constructed duality-symmetric D3-brane action also
has symmetries such as diffeomorphism and kappa symmetry.
The action also gives required equations of motion.
Hamiltonian analysis is also studied.
The decoupling between physical and unphysical sectors
are also shown at the Hamiltonian level.}

\thispagestyle{empty}
\newpage
\tableofcontents

\section{Introduction}
Maxwell's equations in free space are invariant under the transformation $\vec{E}\mapsto\vec{B}$, $\vec{B}\mapsto-\vec{E}$. More generally, it is invariant under the $SO(2)$ rotation of the doublet $(\vec{E},\vec{B})$.
This is an example of duality symmetry.
Extensions to theories with $n$ gauge fields with duality symmetry can also be made.
Couplings to other fields can also be included which in some case can enhance duality symmetry group. These extensions are studied in supergravity and string theory. See for example \cite{Gaillard:1981rj}, \cite{Gibbons:1995cv}, \cite{Ceresole:1995jg}, \cite{Gibbons:1995ap}, \cite{Gaillard:1997rt}, \cite{Andrianopoli:1996bq}, \cite{Andrianopoli:1996ve}, \cite{Gaillard:1997zr}, \cite{Araki:1998nn}, \cite{Aschieri:2008ns}.

For the standard Maxwell theory,
duality symmetry is not a symmetry at the level of the action.
It is possible to construct a theory giving rise to Maxwell equations in free space
and such that the action is duality-symmetric.
To do this, one introduces another gauge field into the theory.
The equations of motion should give the 
constitutive relations
which algebraically (no derivative on the field strengths)
relate the field strengths of the two gauge fields.

Early constructions of duality-symmetric actions are given in \cite{Zwanziger:1970hk}, \cite{Deser:1976iy}.
In these constructions, manifest Lorentz invariant are given up.
The actions are still Lorentz invariant but the form of the Lorentz transformations are modified.

It is possible to restore manifest Lorentz symmetry.
This is by using the PST formalism \cite{Pasti:1995ii}, \cite{Pasti:1995tn}.
One introduces an auxiliary scalar field,
which has no dynamics and can be gauge-fixed to recover the non-manifest Lorentz invariant theory.
It can be realised \cite{Maznytsia:1998xw} by gauge fixing the auxiliary scalar field that the manifestly Lorentz invariant action is reduced to the action of 
\cite{Deser:1976iy}.
An alternative manifestly Lorentz invariant action can be obtained by dualising the auxiliary field, giving rise to the action in the dual formulation.
Under gauge fixing of the auxiliary field, the action in the dual formulation is reduced in the action of \cite{Zwanziger:1970hk}.
Further developments are given for example in \cite{Bekaert:2001wa}, \cite{Bossard:2011ij}, \cite{Pasti:2012wv}, \cite{Lee:2013ewa}.
An advantage of the PST formalism is that coupling to gravity can straightforwardly be introduced.

Duality-symmetric Maxwell theory can also be extended.
In this paper, we are particularly interested in the extension
to describe supersymmetric D3-brane action in the Green-Schwarz formalism,
i.e. the bosonic D3-brane worldvolume is embedded in the ten-dimensional
type IIB supergravity background.

In the literature, the first construction of duality-symmetric D3-brane action
coupled to the background type IIB supergravity is presented in \cite{Cederwall:1997ab}.
In this formalism, which genaralises the manifestly duality-invariant superstring action \cite{Cederwall:1997ts},
the action is covariant and is manifestly duality-invariant.
The supergravity background is also written in the form which is manifestly duality-invariant.
The kappa-symmetry is also explicitly shown.
The constructed D3-brane action
is consistent with the twisted self-duality condition
which is imposed at the level of the equations of motion.

Within the PST formalism, the duality-symmetric D3-brane actions are constructed in \cite{Berman:1997iz}, \cite{Nurmagambetov:1998gp}, \cite{Suzuki:1999aa}.
The twisted self-duality condition is obtained as an equation of motion.
The action is also manifestly diffeomorphism invariant and is shown to have kappa-symmetry.
One may obtain an action in the dual formulation \cite{Vanichchapongjaroen:2017zuh}, in the sense of dualisation of the auxiliary scalar field. The dualised action is also manifestly diffeomorphism invariant and produces twisted self-duality condition as an equation of motion.
However, the dualised action has three auxiliary scalar fields instead of just one auxiliary scalar field required in the original action. At the linearised level with background fields turned off and after the gauge-fixing of the auxiliary scalars, the action of \cite{Berman:1997iz}, \cite{Nurmagambetov:1998gp}, \cite{Suzuki:1999aa} reduces to the action of \cite{Deser:1976iy}, whereas the action of \cite{Vanichchapongjaroen:2017zuh} reduces to the action of
\cite{Zwanziger:1970hk}.

A recent alternative formalism
is also successful in constructing
duality symmetric actions
\cite{Avetisyan:2021heg},
\cite{Mkrtchyan:2022xrm}.
A key advantage of this formalism is that
the auxiliary scalar field appears polynomially in the actions
which is in contrast to the PST formalism
in which the auxiliary scalar field appears in a non-polynomial way.

Another recent alternative formalism is the Sen formalism \cite{Sen:2015nph}, \cite{Sen:2019qit}.
It is originated from the construction of a covariant type IIB supergravity action
inspired from the Sen superstring field theory \cite{Sen:2015uaa}.
The construction can naturally be applied to chiral $(2p)$-form fields in $4p+2$ dimensional spacetime.
A key feature of the Sen formalism is that the chiral form fields
are not independent fields. Instead, they are composed from independent fields
which have non-standard diffeomorphism symmetry.
Another important key feature is that
a $(2p+1)$-form field is self-dual on-shell
and is closed off-shell.
See \cite{Lambert:2019diy}, \cite{Andriolo:2020ykk}, \cite{Vanichchapongjaroen:2020wza}, \cite{Evnin:2022kqn}, \cite{Andrianopoli:2022bzr}, \cite{Lambert:2023qgs}, \cite{Phonchantuek:2023iao}, \cite{Chakrabarti:2023czz}, \cite{Hull:2023dgp} for some developments and extensions.

It is apparent that
the four-dimensional duality-symmetric theories in the PST formalism
have the form that are resemblance to the
theory of $4p+2$ dimensional chiral $(2p)$-form fields.
For example, one may construct a duality-symmetric D3-brane action from the dimensional reduction \cite{Berman:1997iz}, \cite{Nurmagambetov:1998gp} of an M5-brane action.
The resemblance should also be expected in the Sen formalism.
The reference \cite{Sen:2019qit} suggests, by considering a dimensional reduction of a six-dimensional chiral $2$-form theory to a four-dimensional duality-symmetric theory, that this should be the case.
See also \cite{Aggarwal:2025fiq} which directly writes down a four-dimensional duality symmetric theory in the Sen formalism and studies quantum properties.
In the literature, there is still no extension of the duality-symmetric theory in the Sen formalism
giving rise to non-linear constitutive relations.

A main goal of this paper is therefore to construct a duality-symmetric D3-brane action in the Sen formalism. In passing, we construct a duality-symmetric action with non-linear twisted self-duality conditions
(which are related to non-linear constitutive relations).

The rest of this paper is organised as follows.
In section \ref{sec:review},
we give brief reviews on duality-symmetric theories
and on the standard D3-brane action.
This is to set the scene and to introduce conventions to be used in this paper.
In section \ref{sec:result}, we derive and present key results of this paper.
We will first describe how the Sen formalism can be used to construct $SL(2,\RR)$
duality-symmetric actions. A key ingredient of the construction
of each action is a composite field
which satisfies
an off-shell non-linear twisted self-duality condition
which is equivalent to a constitutive relation.
The non-linear twisted self-duality condition
can then be used to construct a Sen action
which gives the required equations of motion.
In fact the action also contains coupling to external scalar fields,
a doublet of $2$-form fields,
and two metrics.
Then after appropriately specifying the potential term,
identifying external fields with appropriate pull-backs of supergravity background superfields,
and adding the remaining part of Wess-Zumino term,
we obtain a duality-symmetric D3-brane action in the Sen formalism. The action also has kappa-symmetry.
In section \ref{sec:properties}, we show that the actions constructed in section \ref{sec:result}
have $SL(2,\RR)$ duality symmetry and diffeomorphism symmetry.
We also present Hamiltonian constraint analysis.
In section \ref{sec:conclusion}, we conclude this paper and propose future works.

\section{Reviews and conventions}\label{sec:review}

\subsection{Duality-symmetric theories}
We work in four dimensional spacetime with signature $(-+++)$.
The coordinates are labelled using Greek indices, for example $x^\m$, $\m = 0,1,2,3$.
The wedge product of all coordinate basis one-form is
\be
dx^{\m}\w dx^{\n}\w dx^{\r}\w dx^{\s}
=\e^{\m\n\r\s}d^4 x,
\ee
where $\e^{\m\n\r\s}$ is the Levi-Civita symbol with $\e^{0123} = -\e_{0123} = 1$.
Differential forms are expressed in coordinate basis as
\be
B_{(p)}
=\ove{p!}dx^{\m_1}\w\cdots\w dx^{\m_p}B_{\m_{p}\cdots\m_1}.
\ee
The exterior derivative and interior products are defined to act from the right.

Consider a system
of an Abelian gauge field $A$
interacting with external metric field $g_{\m\n}$.
The Lagrangian $\cL$ depends on $f = dA$ and $g_{\m\n}$,
but not on derivatives of $f$.
Define the dual field strength $\hat{f}$ from
\be
\d_{f}\cL
=\hat{f}\w\d f,
\ee
or
\be\label{duality_relation}
(\bullet \hat{f})^{\m\n}
=-\frac{2}{\sqrt{-g}}\frac{\pa\cL}{\pa f_{\m\n}},
\ee
where the Hodge star operator corresponding to the metric $g$
is defined as
\be
\bullet(dx^{\m_1}\w\cdots\w dx^{\m_p})
=(-1)^{\frac{p(p+1)}{2}}\frac{\sqrt{-g}}{(4-p)!}dx^{\n_{p+1}}\w\cdots\w dx^{\n_4}\e_{\r_1\cdots\r_p\n_{p+1}\cdots\n_4}g^{\m_1\r_1}\cdots g^{\m_p\r_p}.
\ee
The condition \eq{duality_relation} is called the constitutive relation.

The equation of motion for the Lagrangian $\cL$
can be written as $d\hat{f} = 0$,
whereas the Bianchi identity is $df = 0$.
One is interested in theories with duality symmetry.
For these theories,
duality transformations on $(f, \hat{f})$
should be
such that the system of the
equation of motion, Bianchi identity, and the constitutive relation is invariant.
In order for the theory with Lagrangian $\cL$ to be on-shell symmetric under the $SO(2)$ duality rotations
\be
\begin{pmatrix}
f' &
\hat{f}'
\end{pmatrix}
=
\begin{pmatrix}
f &
\hat{f}
\end{pmatrix}
\begin{pmatrix}
\cos\th & \sin\th\\
-\sin\th & \cos\th
\end{pmatrix}
\ee
the Gaillard-Zumino condition \cite{Gaillard:1981rj}, \cite{Gibbons:1995ap}, \cite{Gaillard:1997zr}
\be\label{GZcond}
\hat{f}_{\m\n}\hat{f}_{\r\s}\e^{\m\n\r\s}
+f_{\m\n}f_{\r\s}\e^{\m\n\r\s}
=0
\ee
should be satisfied.

A pair of scalar fields, i.e. an axion $C_0$ and a dilaton $\f$, can be introduced into the theory
to enhance the $SO(2)$ symmetry to $Sp(2,\RR)$.
The Gaillard-Zumino condition is also generalised.
Let $\t = C_0 + ie^{-\f}$.
Consider a Lagrangian $\cL(f_{\m\n},\t,g_{\m\n})$, where $f = dA$.
Define $\hat{f}$ in the same way as eq.\eq{duality_relation}.
It can be shown (see for example \cite{Gaillard:1997rt}) that in order for the $SL(2,\RR)\cong Sp(2,\RR)$ transformation (with $A,B,C,D\in\RR$ and that $AD - BC = 1$)
\be
\begin{pmatrix}
f' &
\hat{f}'
\end{pmatrix}
=
\begin{pmatrix}
f &
\hat{f}
\end{pmatrix}
\begin{pmatrix}
A & C\\
B & D
\end{pmatrix},\qquad
\t' = \frac{D\t + C}{B\t + A}
\ee
to transform
\be
d\hat{f} = 0,\qquad df = 0,\qquad
(\bullet \hat{f})^{\m\n}
=-\frac{2}{\sqrt{-g}}\frac{\pa\cL(f_{\m\n},\t,g_{\m\n})}{\pa f_{\m\n}}
\ee
into
\be
d\hat{f}' = 0,\qquad df' = 0,\qquad
(\bullet \hat{f}')^{\m\n}
=-\frac{2}{\sqrt{-g}}\frac{\pa\cL(f'_{\m\n},\t',g_{\m\n})}{\pa f'_{\m\n}},
\ee
the generalised Gaillard-Zumino condition
\be\label{Gen-GZ}
(e^{-\f} + C_0^2 e^\f)f_{\m\n}f_{\r\s}\e^{\m\n\r\s}
-2C_0 e^\f f_{\m\n}\hat{f}_{\r\s}\e^{\m\n\r\s}
+e^\f\hat{f}_{\m\n}\hat{f}_{\r\s}\e^{\m\n\r\s}
=0
\ee
should be satisfied.

Given a theory of one gauge field $A$ with on-shell duality symmetry,
its action is not duality invariant.
It is possible to construct an alternative action
which is duality invariant.
For this, one defines another gauge field $\hat{A}$ such that $\hat{f} = d\hat{A}$.
An action for $A$ and $\hat{A}$ can be constructed
such that it is invariant under the duality transformations
and gives the constitutive relations \eq{duality_relation} as equations of motion.
This kind of action is called duality-symmetric action.

One may collect $A\equiv A^1$ and $\hat{A}\equiv A^{\hat{1}}$ into a $2$-tuple
as
\be
(A^I) \equiv (A^1,A^{\hat{1}}),
\ee
where upper-case Latin indices label the position of $2$-tuple,
for example $I = 1,\hat{1}$.
Denote $\vec{u}_1 = (1,0)$ and $\vec{u}_{\hat{1}} = (0,1)$.
We may use vector notation to describe 2-tuple,
for example
\be
\vec{A}
\equiv A^I \vec{u}_I.
\ee
Similarly,
\be
\vec{f}
\equiv f^I \vec{u}_I.
\ee
One will also require $2$-tuple $p$-forms:
\be
\vec{B}_{(p)}
\equiv B_{(p)}^I\vec{u}_I.
\ee

Consider introducing
a collection $\F$ of scalar fields
into the system.
The duality group $G\subset Sp(2,\RR)$ is the group of 
the duality transformations which act linearly on the vector space of $2$-tuples.
The scalar fields $\F$ parametrise a coset space $G/H$,
where $H$ is the maximal compact subgroup of $G$.
The $\F$-dependent invertible matrix
$M_{IJ} = M(\vec{u}_I,\vec{u}_J)
=M(\vec{u}_J,\vec{u}_I)$
can be constructed.
Here $M$ is a bilinear map on the vector space of $2$-tuples.
One may also define a symplectic form
$\Omega$ such that
$
\Omega_{IJ} = \Omega(\vec{u}_I,\vec{u}_J)
=-\Omega(\vec{u}_J,\vec{u}_I).
$
In particular, the components are defined as
$\Omega_{IJ} = \e_{IJ}$,
which is a Levi-Civita symbol with $\e_{1\hat{1}} = 1$.
One may also define $\Omega^{IJ} = -\Omega^{JI}$
such that $\Omega^{IJ}\Omega_{JK} = -\d^I_K$.
The map $M$ should also satisfy $\Omega_{IJ} = M_{IK}\Omega^{KL}M_{LJ}$.

In this paper,
we will only consider the case where $\F$ is an axion-dilaton pair $\t = C_0 + ie^{-\f}$.
In this case, the matrix $M_{IJ}$ is given by
\be\label{M_mat}
(M_{IJ})
=
\begin{pmatrix}
e^{-\f} + C_0^2 e^\f & -C_0 e^\f\\
-C_0 e^\f & e^\f
\end{pmatrix}
.
\ee

One may then define a linear map $J$ which sends a $2$-tuple
to a $2$-tuple such that $J\vec{u}_I = J^J{}_I\vec{u}_J$
and that
\be
J^J{}_I = \Omega^{JK}M_{KI},\qquad
J^J{}_K J^K{}_I = -\d^J_I.
\ee
Explicitly,
\be\label{J_mat}
(J^J{}_I)
=
\begin{pmatrix}
-C_0 e^\f & e^\f\\
-e^{-\f} - C_0^2 e^\f & C_0 e^\f
\end{pmatrix}
.
\ee
The properties can be summarised
using index-free notations as
\be\label{OMJ}
\Omega(\vec{v},J\vec{w})=-M(\vec{v},\vec{w}),\qquad
M(J\vec{v},J\vec{w}) = M(\vec{v},\vec{w}),\qquad
\Omega(J\vec{v},J\vec{w})
=\Omega(\vec{v},\vec{w}),
\ee
where $\vec{v}$ and $\vec{w}$ are any $2$-tuples.
Furthermore, $J^2 = -1$.
Note that $M_{IJ}$ and $J^I{}_J$ depend on $\F$, whereas
$\Omega_{IJ}$ is constant.

Let us define the operation
\be
*\equiv J\bullet,
\ee
which acts on $2$-tuple differential forms as
\be
*\vec{A}_{(p)}
=(\bullet A^I_{(p)}) J^J{}_I\vec{u}_J.
\ee
The $*$ operation satisfy $** = 1$
when applied on any $2$-tuple $2$-form.
Wedge product between $2$-tuple differential forms
are defined as
\be
\vec{A}_{(p)}\w\vec{B}_{(q)}
\equiv A^I_{(p)}\w B^J_{(q)}\vec{u}_I\otimes\vec{u}_J.
\ee
The map $\Omega$, being a bilinear map on the vector space of $2$-tuples can also be defined as\footnote{A more precise mathematical prescription is by using the universal property that $\Omega$, a bilinear map on the vector space of $2$-tuples, induces a linear map $\hat{\Omega}$ on the tensor product of two vector spaces of $2$-tuples. However, for the purpose of this paper, we do not use different notations to distinguish these two related maps. Furthermore, the definition in eq.\eq{Omega_on_tensor} should be sufficient.} a linear map on the tensor product of two vector spaces of $2$-tuples such that
\be\label{Omega_on_tensor}
\Omega(\a^{IJ}\vec{u}_I\otimes\vec{u}_J)
= \a^{IJ}\Omega(\vec{u}_I\otimes\vec{u}_J)
= \a^{IJ}\Omega(\vec{u}_I,\vec{u}_J)
= \a^{IJ}\Omega_{IJ}.
\ee
So for example,
\be
\Omega(\vec{A}_{(p)}\w\vec{B}_{(q)})
=A^I_{(p)}\w B^J_{(q)}\Omega_{IJ}.
\ee
Similarly, we may also define $M$ as a linear map on the tensor product of two vector spaces of $2$-tuples
by
\be
M(\a^{IJ}\vec{u}_I\otimes\vec{u}_J)
= \a^{IJ}M(\vec{u}_I\otimes\vec{u}_J)
= \a^{IJ}M(\vec{u}_I,\vec{u}_J)
= \a^{IJ}M_{IJ}.
\ee
Useful identities are 
\be
\Omega(\vec{A}_{(p)}\w\vec{B}_{(q)})
=-(-1)^{pq}\Omega(\vec{B}_{(q)}\w\vec{A}_{(p)}),
\ee
\be
\Omega(\vec{A}_{(p)}\w*\vec{B}_{(q)})
=\Omega(\vec{B}_{(q)}\w*\vec{A}_{(p)}),
\ee
\be
\Omega(\vec{A}_{(p)}\w *\vec{B}_{(q)})
=-M(\vec{A}_{(p)}\w\bullet\vec{B}_{(q)}).
\ee

As an explicit example of a theory with duality symmetry,
consider a system
in which a gauge field $A_\m$ is coupled to background fields $g_{\m\n}, \t = C_0 + ie^{-\f}, \vec{X}_{\m\n} = (X_{\m\n},\hat{X}_{\m\n})$.
We will denote the field strength of $A_\m$
as $f_{\m\n} = \pa_\m A_\n - \pa_\n A_\m - X_{\m\n}$.
By convention, this field strength is not simply the exterior derivative of the gauge field.
The appearance of $X_{\m\n}$ in the definition of field strength is analogous to that of the pull-back of background NS-NS B-field to D-brane worldvolume, see for example eq.\eq{fdAB}.
The action of the particular system of interest is
\be\label{1field_Mtau}
S = \int d^4 x\lrbrk{-\ove{4}\sqrt{-g}e^{-\f}f_{\m\n}f^{\m\n} + \ove{8}C_0 f_{\m\n}\e^{\m\n\r\s}f_{\r\s} + \ove{4} f_{\m\n}\e^{\m\n\r\s}\hat{X}_{\r\s}}.
\ee
Define
\be\label{bullet_fhat}
(\bullet\hat{f})^{\m\n}\equiv
-\frac{2}{\sqrt{-g}}\frac{\d S}{\d f_{\m\n}} - (\bullet \hat{X})^{\m\n},
\ee
which gives
\be\label{lin_consti}
(\bullet\hat{f})^{\m\n}=e^{-\f}f^{\m\n} + C_0(\bullet f)^{\m\n}.
\ee
The Bianchi identity and equation of motion for $A_\m$ are
\be\label{EOM_Bian}
\pa_{[\m}(f+X)_{\n\r]} = 0,\qquad
\pa_{[\m}(\hat{f}+\hat{X})_{\n\r]} = 0.
\ee
Note that although
the above equations seem to contain $X_{\m\n}, \hat{X}_{\m\n}$
and the other background fields,
these equations do not restrict the form of background fields.
Due to the definition of $f_{\m\n}$,
the first equality of eq.\eq{EOM_Bian} is simply $d^2 A= 0$,
which is trivial due to the identity $d^2 = 0$.
As for the second equality of eq.\eq{EOM_Bian},
it can also be expressed as
$0 = d(-e^{-\f}\bullet (dA - X) + C_0 (dA - X)  + \hat{X})$,
which describes the dynamics of $A_\m$
in the presence of background fields $g_{\m\n}, \f, C_0, X_{\m\n},\hat{X}_{\m\n}$.
The system described by the action \eq{1field_Mtau} is duality-symmetric.
This means that
under $SL(2,\RR)$ duality transformations
\be
\begin{pmatrix}
f' &
\hat{f}'
\end{pmatrix}
=
\begin{pmatrix}
f &
\hat{f}
\end{pmatrix}
\begin{pmatrix}
A & C\\
B & D
\end{pmatrix}
,\qquad
\begin{pmatrix}
X' &
\hat{X}'
\end{pmatrix}
=
\begin{pmatrix}
X &
\hat{X}
\end{pmatrix}
\begin{pmatrix}
A & C\\
B & D
\end{pmatrix}
,\qquad
\t' = \frac{D\t + C}{B\t + A}
\ee
with parameters $A,B,C,D\in\RR$ such that $AD - BC = 1$,
the set of equations \eq{lin_consti}-\eq{EOM_Bian}
is invariant.

The system described by the action \eq{1field_Mtau}
can also be described by
a duality-symmetric action
in the PST formalism \cite{Pasti:1995ii}, \cite{Pasti:1995tn}:
\be
S = -\ove{4}\int\Omega(\vec{f}\w\cP(1-*)\vec{f} + \vec{f}\w\vec{X}),
\ee
where $\vec{A}$ is the independent field,
$\vec{f} = d\vec{A} - \vec{X}$ is the field strength,
$a$ is an auxiliary scalar field, and
\be
\cP\equiv \frac{g^{-1}da\otimes da}{g^{-1}(da,da)}.
\ee
This action is manifestly duality and diffeomorphism invariant.
It also has the symmetries called PST1 and PST2 symmetries.
The PST1 symmetry is
\be
\d a = 0,\qquad
\d\vec{A} = da\vec{\l},
\ee
for any $2$-tuple $0$-form $\vec{\l}$.
This symmetry is used to realise that the equation of motion of $\vec{A}$
is equivalent to twisted self-duality condition
\be
\vec{f} = *\vec{f},
\ee
which is equivalent to the constitutive relation \eq{lin_consti}.
The PST2 symmetry is
\be
\d \vec{A}
=\frac{\d a}{g^{-1}(u,u)}i_{g^{-1}u}(1-*)\vec{f},
\ee
which can be used to gauge fix $a$
and corresponds to the fact that $a$ is auxiliary field
due to its equation of motion being implied by the equations of motion of $\vec{A}$.

\subsection{The standard D3-brane action}\label{sec:standardD3}
In this subsection, we review standard D3-brane action in the Green-Schwarz formalism. A D3-brane is described as a theory in four-dimensional worldvolume containing a gauge field $A$. The D3-brane is embedded in a ten-dimensional target superspace described by type IIB supergravity.

The standard D3-brane action is however not duality-symmetric.
It is possible to write duality-symmetric D3-brane actions
in the PST formalism \cite{Pasti:1995ii}, \cite{Pasti:1995tn}, \cite{Berman:1997iz}, \cite{Nurmagambetov:1998gp}, \cite{Suzuki:1999aa}, \cite{Vanichchapongjaroen:2017zuh}.
Alternatively, as to be presented in subsection \ref{subsec:SenD3}
as part of the results of this paper,
the duality-symmetric D3-brane action
in the Sen formalism can also be written.

We first review relevant ingredients from type IIB supergravity.
The convention largely agrees with \cite{Cederwall:1996ri}.

Coordinates on the target superspace are $(Z^\uM) = (X^\unm,\th)$,
where $X^{\unm}$ are $10$ bosonic coordinates,
and $\th$ are $32$ fermionic coordinates, which
can be collected as a doublet of Majorana-Weyl spinors of positive chirality.
Indices corresponding to the target space are denoted using underline.

The field content of type IIB supergravity consists of supervielbeins $E^\uA$, a dilaton superfield $\f$, an NS-NS super $B$-field $B_2$, and RR super $C$-fields $C_0, C_2, C_4$.
The supervielbeins are $(E^\uA) = (E^\una, E^\ua)$, where
\be
E^\una = dZ^\uM E^\una_\uM,\qquad
E^\ua = dZ^\uM E^\ua_\uM.
\ee
The index $\una\in\{0,1,2,\cdots,9\}$ is a bosonic tangent space index,
whereas the index $\ua$ is a fermionic tangent space index.
Differential $p$-form superfields are expressed as
\be
C_p = \ove{p!}dZ^{\uM_1}dZ^{\uM_2}\cdots dZ^{\uM_p}C_{\uM_p\cdots \uM_2\uM_1},
\ee
where wedge product notation $\wedge$ is omitted.
The NS-NS super $B$-field has field strength $H_3 = dB_2$.
The RR super $C$-fields have field strengths
\be
R_1 = dC_0,\qquad
R_3 = dC_2 - H_3C_0,\qquad
R_5 = dC_4 - H_3C_2.
\ee
The type IIB supergravity is invariant under supergauge transformation
\be
\d(E^\una\otimes E^\unb)\h_{\una\unb} = 0,\qquad
\d\f = 0,\qquad
\d B_2 = d\l_1,
\ee
\be
\d C_0 = 0,\qquad
\d C_2 = d\m_1,\qquad
\d C_4 = d\m_3 + B_2 d\m_1,
\ee
where $\h_{\una\unb} = \diag(-1,1,\cdots,1)$ is the 10D Minkowski metric
for the bosonic tangent space.

Gamma matrices are expressed as
\be
\G^\una\otimes\s,
\ee
where $\G^\una$ acts on a Majorana-Weyl index,
while $\s$ acts on the doublet index.
The matrices $\s$ are real $2\times 2$ matrices
which are for example
\be
\mathbbm{1}
=
\begin{pmatrix}
1 & 0\\
0 & 1
\end{pmatrix}
,\qquad
i\t_2
=
\begin{pmatrix}
0 & 1\\
-1 & 0
\end{pmatrix}
,\qquad
\t_1
=
\begin{pmatrix}
0 & 1\\
1 & 0
\end{pmatrix}
,\qquad
\t_3
=
\begin{pmatrix}
1 & 0\\
0 & -1
\end{pmatrix}
.
\ee
For convenient, we will omit $\mathbbm{1}$ and $\otimes$.

In the Green-Schwarz formalism,
the D3-brane action encodes the dynamics
of the worldvolume field coupled
to the external target space superfields.
That is, the dynamics of external target space superfields
are not encoded into the D3-brane action
but are pre-determined by the equations of motion and constraints.
One only needs a part of the type IIB supergravity constraints
to prove that the D3-brane action has kappa symmetry.
The relevant constraints consist of
the torsion constraint
\be\label{tcon}
dE^\una + E^\unb\o_\unb{}^\una = iE^\ua\G^\una_{\ua\ub}E^\ub,
\ee
where $\o_\unb{}^\una$ is the connection one-form.
There are also the following constraints on the field strengths:
\be\label{fscons}
\begin{split}
H_3 &= -ie^{\f/2} E^\unc E^\ub E^\ua(\G_\unc\t_3)_{\ua\ub}
+\hlf e^{\f/2} E^\unc E^\unb  E^\ua(\G_{\unb\unc}\t_3\L)_\ua+\cdots,\\
R_1 &= 2e^{-\f}E^\ua(i\t_2\L)_\ua+\cdots,\\
R_3 &= ie^{-\f/2}E^\unc E^\ub E^\ua(\G_\unc \t_1)_{\ua\ub} + \hlf e^{-\f/2} E^\unc E^\unb E^\ua (\G_{\unb\unc}\t_1\L)_\ua + \cdots,\\
R_5 &= \frac{i}{6}E^\ub E^\ua E^\unc E^\unb E^\una (\G_{\una\unb\unc}i\t_2)_{\ua\ub}+\cdots,
\end{split}
\ee
where $\cdots$ are terms which are irrelevant to kappa-symmetry variation,
and
\be
\L_\ua = \hlf E^\uM_\ua\pa_\uM\f.
\ee

The D3-brane is coupled to the background target space superfields
by using pull-backs. For convenient, we will not introduce extra notations to distinguish the background fields and their pull-backs.
The field strength of the gauge field $A$ on the D3-brane
is
\be\label{fdAB}
f = dA - B,
\ee
where $B$ is the pull-back of the NS-NS super $B$-field.

The standard D3-brane action is
\be\label{D3action}
S_{D3}
=-\int d^4 x\sqrt{-\det(g_{\m\n} + e^{-\f/2}f_{\m\n})}
+\int (C_4 + f\w C_2 + \hlf f\w f\w C_0),
\ee
where $g_{\m\n} = E^\una_\m E^\unb_\n \h_{\una\unb},$ and $E^\una_\m = E^\una_\uM \pa_\m Z^\uM.$
Since this action is in the Green-Schwarz formalism,
one should check that it is
invariant under the kappa-symmetry transformation, which is a fermionic transformation ensuring that the number of degrees of freedom for fermions and bosons are matched.
The variation should be applied to the worldvolume field as well as the (pull-back of) the background superfields
and should be consistent with the constraints in eq.\eq{tcon}-\eq{fscons}.
By setting 
\be
\d_\k f = - i_\k H_3,
\ee
where $\k$ is a Majorana-Weyl spinor,
one obtains
\be\label{dkWZ}
\d_\k\int \lrbrk{C_4 + f\w C_2 + \hlf f\w f\w C_0}
=\int\lrbrk{i_\k R_5 + f\w i_\k R_3 + \hlf f\w f\w i_\k R_1},
\ee
where $\d_k = di_\k + i_\k d$ when acting on pull-back of background superfields.
The equation \eq{dkWZ} suggests that if we consider the full action, we only need $\d_\k g_{\m\n}, \d_\k\f ,i_\k R_1, i_\k R_3, i_\k H_3, i_\k R_5$
to obtain the kappa-symmetry variation.
After setting
\be\label{ikE-1}
i_\k E^{\una} = \d_\k Z^{\uM}E^{\una}_{\uM} = 0,
\ee
one obtains the following
\be\label{dkg}
\d_\k g_{\m\n} = 4iE^\ua_{(\m}(\g_{\n)})_{\ua\ub}i_\k E^\ub,\qquad
\d_\k\f = 2\L^\ua (i_\k E)_\ua,
\ee
\be
i_\k R_1 = -2ie^{-\f}\bar{\L}\t_2 i_\k E,
\ee
\be
i_\k R_3 = dx^{\m}\w dx^\n\lrbrk{-2ie^{-\f/2}\bar{E}_\m\g_\n \t_1 + \hlf e^{-\f/2}\bar{\L}\g_{\m\n}\t_1}i_\k E,
\ee
\be
i_\k H_3 = dx^{\m}\w dx^\n\lrbrk{2ie^{\f/2}\bar{E}_\m\g_\n \t_3 + \hlf e^{\f/2}\bar{\L}\g_{\m\n}\t_3}i_\k E,
\ee
\be\label{ikR5}
i_\k R_5 = \ove{3} d^4 x\e^{\m\n\r\s}\bar{E}_\s\g_{\m\n\r}\t_2 i_\k E,
\ee
where
$\g_\m = E^\una_\m\G_\una.$
One then obtains
\be\label{dkSD3}
\begin{split}
\d_\k S_{D3}
&=\int d^4 x\sqrt{-g} 2i\bar{E}_\m (-\g^\m\bar{\g}\t_2 + e^{-\f/2}\tilde{f}^{\m\n}\g_\n\t_1)(1-\bar{\G}_{D3})i_\k E\\
&\qquad-\int d^4 x\frac{\sqrt{-g}}{2}e^{-\f/2} \bar{\L} (\tilde{f}^{\m\n}\g_{\m\n}\t_1
+ i e^{-\f}\tilde{f}^{\m\n}f_{\m\n}\t_2)(1-\bar{\G}_{D3})i_\k E,
\end{split}
\ee
where
\be
i_\k E^{\ua} = \d_\k Z^{\uM}E^{\ua}_{\uM},
\ee
\be
\tilde{f}^{\m\n}
=\hlf\frac{\e^{\m\n\r\s}}{\sqrt{-g}}f_{\r\s},
\ee
\be
\bar\g = -\frac{i}{4!}\frac{\e^{\m\n\r\s}}{\sqrt{-g}}\g_{\m\n\r\s},
\ee
and the matrix $\bar{\G}_{D3}$ is given by \cite{Cederwall:1996ri}, \cite{Cederwall:1996pv}
\be
\sqrt{1-\hlf e^{-\f}[f^2]-\ove{16}e^{-2\f}[f\tilde f]^2}\ \bar{\G}_{D3}
=\bar\g\t_2
-\hlf\tilde f^{\m\n}\g_{\m\n}\t_1
+\frac{i}{4}[f\tilde f]\t_2.
\ee
The matrix $\bar{\G}_{D3}$ satisfies $\bar{\G}_{D3}^2 = 1, \tr(\bar{\G}_{D3}) = 0.$
With these properties, $(1\pm\bar{\G}_{D3})/2$ are projectors of rank $8$ and satisfy
\be
\frac{1\pm\bar{\G}_{D3}}{2}\frac{1\pm\bar{\G}_{D3}}{2} = \frac{1\pm\bar{\G}_{D3}}{2},
\ee
\be
\frac{1\pm\bar{\G}_{D3}}{2}\frac{1\mp\bar{\G}_{D3}}{2} = 0.
\ee
In order for the variation \eq{dkSD3} to vanish one should set
\be
i_\k E^{\ua} = ((1+\bar\G_{D3})\k)^\ua.
\ee
Since $\k$ appears in the field variation only through the projector 
$(1+\bar{\G}_{D3})/2$, only half of the components of $\k$
is responsible for the vanishing of the variation of the action.
These components can be used to gauge away the fermionic degrees of freedom
giving the realisation of the matching between bosonic and fermionic degrees of freedom.

Any action constructed in the Green-Schwarz formalism, in which bosonic worldvolume is embedded into the background target superspace, should possess kappa-symmetry. Otherwise the number of the bosonic and fermionic degrees of freedom would not match.
In practice, one should show that
the kappa-symmetry transformation includes
the field variation $i_\k E^{\ua} = ((1+\bar\G)\k)^\ua$,
where the matrix $\bar{\G}$ should be worked out and should satisfy $\bar{\G}^2 =1, \tr(\bar{\G}) = 0$. The matrix $\bar{\G}$ depends on the theory under consideration.
We have seen that for the action \eq{D3action}, the matrix $\bar{\G}$ is given by $\bar{\G}_{D3}$. In subsection \ref{subsec:SenD3}, we will present the duality-symmetric D3-brane action in the Sen formalism. This action has the kappa-symmetry such that the matrix $\bar{\G}$ with all the desired properties can be constructed and is given by $\bar{\G}_{\textrm{Sen-D3}}$ in eq.\eq{barG_SenD3}.

\section{Twisted self-duality and D3-brane in the Sen formalism}\label{sec:result}
In this section,
we focus on the construction of non-linear duality-symmetric actions in the Sen formalism
as well as the duality-symmetric D3-brane action in the Sen formalism.
Duality-symmetry and diffeomorphism invariance of the non-linear actions
will be shown in the next section.

\subsection{Linear twisted self-duality in the Sen formalism}\label{subsec:quad_TwistedSDSen}

In this subsection, we will present the duality-symmetric action in the Sen formalism
such that it describes the same system as that with one-gauge-field action \eq{1field_Mtau}.
This means that
at the level of equations of motion the system is instead described by the dynamics of $\vec{A}\equiv(A^I_\m)\equiv (A_\m, \hat{A}_\m)$ such that $\vec{f} = (f,\hat{f}) = d\vec{A} - \vec{X}$ satisfies the constitutive relation \eq{lin_consti} which can be recasted as $\vec{f} = *\vec{f}$.
In fact, there are non-standard features
which are typical for the Sen formalism:
\begin{enumerate}
  \item The system described by the Sen formalism is in fact separated into two sectors, uncoupled to each other. One sector is physical and describes
  the original system, i.e. it describes the fields $A^I_\m$ coupled to the external fields $g_{\m\n}, \t, \vec{X}$. We also call the physical sector as the $g$-sector.
  The other which is the additional sector is unphysical.
  It describes the fields\footnote{In this paper, overbar does not mean complex conjugation. In particular,
  barred fields e.g. $\bar{A}_\m^I, \bar{g}_{\m\n}, \bar{\F}$ are not complex conjugates of their unbarred counterparts.} $\bar{A}^I_\m$ coupled to another metric $\bar{g}_{\m\n}$ and another set $\bar{\F}$ of scalar fields. We also call the unphysical sector as the $\bar{g}$-sector.
  \item At the action level, the fields $A^I_\m, \bar{A}^I_\m$ themselves are not independent fields. Instead, the independent fields are a $2$-tuple $1$-form $\vec{P}$ and a $2$-tuple $2$-form $\vec{Q}$ which is twisted self-dual with respect to $\bar{g}_{\m\n}$. A composite field $\vec{H}$, which is a $2$-tuple $2$-form, is twisted self-dual off-shell and is closed on-shell. Later non-linear generalisation would only directly affect the off-shell twisted self-duality condition.
\end{enumerate}

Define the Hodge star operator corresponding to $\bar{g}_{\m\n}$:
\be
\bar{\bullet}(dx^{\m_1}\w\cdots\w dx^{\m_p})
=(-1)^{\frac{p(p+1)}{2}}\frac{\sqrt{-\bar{g}}}{(4-p)!}dx^{\n_{p+1}}\w\cdots\w dx^{\n_4}\e_{\r_1\cdots\r_p\n_{p+1}\cdots\n_4}\bar{g}^{\m_1\r_1}\cdots \bar{g}^{\m_p\r_p}.
\ee
Furthermore, in this paper we will set for definiteness $\bar{\F}$ to be another axion-dilaton pair $\bar{\t} = \bar{C}_0 + ie^{-\bar{\f}}$. Having specified this,
the corresponding internal metric $\bar{M}_{IJ}$ and the complex structure $\bar{J}^J{}_I$
are defined in a similar way as given in eq.\eq{M_mat}-\eq{J_mat}.
That is,
\be\label{JbMb_mat}
(\bar{M}_{IJ})
=
\begin{pmatrix}
e^{-\bar{\f}} + \bar{C}_0^2 e^{\bar{\f}} & -\bar{C}_0 e^{\bar{\f}}\\
-\bar{C}_0 e^{\bar{\f}} & e^{\bar{\f}}
\end{pmatrix}
,\qquad
(\bar{J}^J{}_I)
=
\begin{pmatrix}
-\bar{C}_0 e^{\bar{\f}} & e^{\bar{\f}}\\
-e^{-\bar{\f}} - \bar{C}_0^2 e^{\bar{\f}} & \bar{C}_0 e^{\bar{\f}}
\end{pmatrix}
.
\ee
Define the linear maps $\bar{M}$ and $\bar{J}$ such that
$\bar{M}_{IJ} = \bar{M}(\vec{u}_I, \vec{u}_J) = \bar{M}(\vec{u}_J, \vec{u}_I)$
and $\bar{J}\vec{u}_I = \bar{J}^J{}_I\vec{u}_J$.
Properties of $\bar{M}$ and $\bar{J}$ are
\be
\Omega(\vec{v},\bar{J}\vec{w})=-\bar{M}(\vec{v},\vec{w}),\qquad
\bar{M}(\bar{J}\vec{v},\bar{J}\vec{w}) = \bar{M}(\vec{v},\vec{w}),\qquad
\Omega(\bar{J}\vec{v},\bar{J}\vec{w})
=\Omega(\vec{v},\vec{w}),
\ee
where $\vec{v}$ and $\vec{w}$ are any $2$-tuples.
Furthermore, $\bar{J}^2 = -1$.

Similar to the case of chiral form theories,
the independent fields of a duality-symmetric Lagrangian in the Sen formalism
are not $A^I_\m, \bar{A}^I_\m$. Instead, they are
a $2$-tuple $1$-form
$\vec{P}$
and 
a $2$-tuple $2$-form $\vec{Q}$
satisfying the twisted self-duality condition with respect to $\bar{\t}, \bar{g}_{\m\n}$:
\be\label{Qsd}
\vec{Q} = \bar{*}\vec{Q},
\ee
where $\bar{*} = \bar{J}\bar{\bullet}$.
We treat scalar fields $\t, \bar{\t}$ and metrics $g, \bar{g}$
as external. Furthermore, we will also include an external $2$-tuple $2$-form source $\vec{X}$.

As a feature of the Sen formalism,
the actions contain
the following composite fields,
\be\label{vecR_defn}
\vec{R} \equiv \vec{R}(\vec{Q},\t,\bar{\t},\vec{X},g,\bar{g}),
\ee
\be\label{vecH_defn}
\vec{H} \equiv \vec{H}(\vec{Q},\t,\bar{\t},\vec{X},g,\bar{g})
= \vec{Q} - \vec{R} - \vec{X}.
\ee
The field $\vec{R}$ is twisted anti-self-dual with respect to $\bar{\t}, \bar{g}_{\m\n}$,
that is
\be\label{vecR_asd}
\vec{R} = -\bar{*}\vec{R}.
\ee
In order to completely specify $\vec{R}$ and $\vec{H}$
in terms of the independent fields and background fields,
twisted self-duality condition for $\vec{H}$ should also be imposed.
For the theory with linear twisted self-duality condition,
one imposes
\be\label{quad_twisted_sd}
\vec{H} = *\vec{H}.
\ee
In order to solve eq.\eq{Qsd}, \eq{vecH_defn}-\eq{quad_twisted_sd},
it is convenient to define
$\vec{X}^{\pm}\equiv (1\pm\bar{*})\vec{X}/2$
and
\be\label{QXRX}
\vec{Q}_X\equiv \vec{Q} - \vec{X}^+,\qquad
\vec{R}_X\equiv \vec{R} + \vec{X}^-.
\ee
Then eq.\eq{Qsd}, \eq{vecH_defn}-\eq{quad_twisted_sd} become
\be\label{QXRX_tsd}
\vec{Q}_X = \bar{*}\vec{Q}_X,\qquad
\vec{R}_X = -\bar{*}\vec{R}_X,
\ee
\be\label{H_tsd}
\vec{H} = \vec{Q}_X - \vec{R}_X,\qquad
\vec{H} = *\vec{H}.
\ee
Define
\be\label{Tdefn}
T\equiv C_0 + e^{-\f} \bullet,\qquad
\bar{T}\equiv \bar{C}_0 + e^{-\bar{\f}}\bar{\bullet}.
\ee
We are applying $T$ and $\bar{T}$ to $2$-forms.
So $\bullet^2 = -1 = \bar{\bullet}^2$,
but in general, $\bullet\bar{\bullet} \neq \bar{\bullet}\bullet$.
Hence $T$ and $\bar{T}$ are considered as non-commuting complex numbers.
Define complex conjugations of $T$ and $\bar{T}$ as
\be\label{Tsdefn}
T^*\equiv C_0 - e^{-\f} \bullet,\qquad
\bar{T}^*\equiv \bar{C}_0 - e^{-\bar{\f}}\bar{\bullet}.
\ee
Denote $\vec{Q}_X\equiv(Q_X,\hat{Q}_X), \vec{R}_X\equiv(R_X,\hat{R}_X),
\vec{H}\equiv(F,\hat{F})$.
From eq.\eq{QXRX_tsd}-\eq{H_tsd}, we have
\be\label{QXTQ}
\hat{Q}_X = \bar{T}^*Q_X,\qquad
\hat{R}_X = \bar{T}R_X,
\ee
\be
F = Q_X - R_X,\qquad
\hat{F} = \hat{Q}_X - \hat{R}_X,\qquad
\hat{F} = T^*F,
\ee
which give
\be\label{RandRhat}
R_X = (\bar{T} - T^*)^{-1}(\bar{T}^* - T^*)Q_X,\qquad
\hat{R}_X = \bar{T}(\bar{T} - T^*)^{-1}(\bar{T}^* - T^*)Q_X,
\ee
\be\label{HandHhat}
F = (\bar{T} - T^*)^{-1}(\bar{T} - \bar{T}^*)Q_X,\qquad
\hat{F} = T^*(\bar{T} - T^*)^{-1}(\bar{T} - \bar{T}^*)Q_X.
\ee

From eq.\eq{RandRhat},
it can be seen that $\vec{R}_X = (R_X,\hat{R}_X)$ is linearly dependent on $Q_X$
(due to self-duality of $\vec{Q}_X$,
only half of the components remains,
which is chosen to be $Q_X$)
but is non-linearly dependent on $\t,\bar{\t},g_{\m\n},\bar{g}_{\m\n}$.
This feature is analogous to the one encountered in the Sen formalism for self-dual $(2p+1)$-forms in $(4p+2)$-dimensional spacetime.
Two metrics are also required in the construction.
In the quadratic theory,
ignoring the external differential form fields
to simplify the discussion, the composite $(2p+1)$-form
$R$ is related to the independent self-dual $(2p+1)$-form field $Q$
by $R = {\cal M} Q$,
where ${\cal M}$ is
a linear map from a $(2p+1)$-form to a $(2p+1)$-form
and is
non-linearly dependent on the two metrics.
See \cite{Sen:2015nph}, \cite{Sen:2019qit}, \cite{Andriolo:2020ykk}, \cite{Vanichchapongjaroen:2020wza} for the forms of ${\cal M}$.

When generalising to theories with non-linear twisted self-duality,
$\vec{R}_X$ would no longer depend linearly on $Q_X$.
As to be seen in eq.\eq{RXQX_nl},
$\vec{R}_X$ would be non-linear in $Q_X$ while its dependence on
$\t,\bar{\t},g_{\m\n},\bar{g}_{\m\n}$ would be even more involved.

The quadratic action for duality-symmetric theory in the Sen formalism has been written down and studied in \cite{Sen:2019qit}, \cite{Aggarwal:2025fiq}. The form of the action resembles the actions for self-dual forms in the Sen formalism \cite{Sen:2015nph}, \cite{Sen:2019qit}.
So it is natural to propose that when written in the notation of this paper,
the quadratic action for duality symmetric theory in the Sen formalism
reads
\be\label{quadratic-tSDSen}
\begin{split}
S_{\textrm{Sen,quadratic}}
&=-\int\Omega\lrbrk{\ove{4}d\vec{P}\w\bar{*}d\vec{P} - \vec{Q}\w d\vec{P} + \hlf\vec{Q}\w\vec{R} - \hlf\vec{H}\w\vec{X}}\\
&=\int\bigg(\ove{4}\bar{M}_{IJ}(\bar{\t})dP^I\w\bar{\bullet}dP^J
+\Omega_{IJ}Q^I\w dP^J\\
&\qquad\qquad-\hlf\Omega_{IJ}Q^I\w R^J
+\hlf\Omega_{IJ}H^I\w X^J\bigg).
\end{split}
\ee
The independent fields are $\vec{P}, \vec{Q}$,
such that $\vec{Q}$ is subject to eq.\eq{Qsd}.
One may also express the action \eq{quadratic-tSDSen}
without the composite fields $\vec{R}, \vec{H}$.
For this, one first writes the action as
\be\label{quadratic-tSDSen-2}
\begin{split}
S_{\textrm{Sen,quadratic}}
&=-\int\Omega\bigg(\ove{4}d\vec{P}\w\bar{*}d\vec{P} - \vec{Q}_X\w d\vec{P} + \hlf\vec{Q}_X\w\vec{R}_X\\
&\qquad\qquad- \vec{X}^+\w d\vec{P} - \vec{Q}_X\w\vec{X}^{-} - \hlf\vec{X}^+\w\vec{X}^-\bigg).
\end{split}
\ee
Then by using eq.\eq{QXRX}, \eq{QXTQ}, \eq{RandRhat}-\eq{HandHhat},
and writing $\vec{P} = (P,\hat{P}), \vec{Q} = (Q,\hat{Q})$,
one obtains
\be\label{quadratic-tSDSen-3}
\begin{split}
S_{\textrm{Sen,quadratic}}
&=\ove{4}\int e^{\bar{\f}} \lrbrk{(e^{-2\bar{\f}} + \bar{C}_0^2) dP\w\bar{\bullet} dP 
-2\bar{C}_0 dP\w\bar{\bullet} d\hat{P} + d\hat{P}\w\bar{\bullet} d\hat{P}}\\
&\qquad+\int Q\w(d\hat{P} - \bar{T}^* dP)
-\int e^{-\bar{\f}}Q\w\bar{\bullet}(\bar{T} - T^*)^{-1}(\bar{T}^* - T^*)Q\\
&\qquad
+2\int e^{-\bar{\f}} Q\w\bar{\bullet}(\bar{T} - T^*)^{-1}(\hat{X} - T^*X)
\\
&\qquad+\hlf\int(\hat{X} - \bar{T}X)\w(\bar{T} - T^*)^{-1}(\hat{X} - T^* X).
\end{split}
\ee
The independent fields
for the action in this form are $P, \hat{P}, Q$.

The action \eq{quadratic-tSDSen} is invariant under the $SL(2,\RR)$ duality transformations\footnote{By the convention of this paper, $2$-tuples are row vectors.}
\be\label{SL2_1}
\vec{P}'
=\vec{P}
\begin{pmatrix}
A & C\\
B & D
\end{pmatrix}
,\qquad
\vec{Q}'
=
\vec{Q}
\begin{pmatrix}
A & C\\
B & D
\end{pmatrix}
,\qquad
\vec{X}'
=
\vec{X}
\begin{pmatrix}
A & C\\
B & D
\end{pmatrix},
\ee
\be\label{SL2_2}
\t' = \frac{D\t + C}{B\t + A},\qquad
\bar{\t}' = \frac{D\bar{\t} + C}{B\bar{\t} + A},
\ee
with parameters $A,B,C,D\in\RR$ such that $AD - BC = 1$.
To see this, it is helpful
to consider the form of the action as given by eq.\eq{quadratic-tSDSen-2}.
It is clear that the only non-trivial check is to show that
\be
\int\Omega(\vec{Q}_X\w\vec{R}_X)
\ee
is duality invariant
since the other terms of eq.\eq{quadratic-tSDSen-2}
are clearly invariant under eq.\eq{SL2_1}-\eq{SL2_2}.

Under the $SL(2,\RR)$ transformations \eq{SL2_1}-\eq{SL2_2}
it can be worked out, with careful consideration of non-commutativity of $T$ and $\bar{T}$, that
$\vec{R}_X = (R_X,\hat{R}_X), \vec{H} = (F,\hat{F})$
given by eq.\eq{RandRhat}-\eq{HandHhat} transform as
\be\label{RH_duality_trafo}
\vec{R}'_X
=
\vec{R}_X
\begin{pmatrix}
A & C\\
B & D
\end{pmatrix}
,\qquad
\vec{H}'
=
\vec{H}
\begin{pmatrix}
A & C\\
B & D
\end{pmatrix}.
\ee
This result suggests that although $\vec{R}_X$ and $\vec{H}$ are composite fields
whose form are non-linear in non-commuting operators $T$ and $\bar{T}$,
the duality transformations of $\vec{R}_X$ and $\vec{H}$ turn out to be the standard linear transformation.

Since both $\vec{Q}_X$ and $\vec{R}_X$ transform linearly under $SL(2,\RR)$ duality transformation, it is then straightforward that $\Omega(\vec{Q}_X\w\vec{R}_X)$ is invariant under the $SL(2,\RR)$ duality transformation.
We have thus shown that the action \eq{quadratic-tSDSen-2}, or equivalently the action \eq{quadratic-tSDSen}, is $SL(2,\RR)$ duality invariant.

It is also important to show that the action \eq{quadratic-tSDSen}
is diffeomorphism invariant.
As is typical in the Sen formalism,
the complete check of diffeomorphism invariance
is non-trivial.
The discussion is postponed to subsection \ref{subsec:diffeo},
in which we discuss and show the diffeomorphism invariance of 
the non-linear generalisation of the action \eq{quadratic-tSDSen}.

Let us now work out and analyse the equations of motion.
For this, one may
either vary the action \eq{quadratic-tSDSen} with respect to $\vec{P}$ and $\vec{Q}$, keeping in mind eq.\eq{Qsd}-\eq{quad_twisted_sd},
or vary the action \eq{quadratic-tSDSen-3} with respect to $P, \hat{P},$ and $Q$.
Both approaches lead to the same result.
It is simpler to follow the former approach.
We first note that
eq.\eq{quad_twisted_sd}
implies
$\Omega(\vec{H}\w\d_Q\vec{H}) = 0,$
which is equivalent to the identity
\be\label{quaddQ}
\d_Q\lrbrk{\hlf\Omega\lrbrk{\vec{Q}\w\vec{R} - \vec{H}\w\vec{X}}}
=\Omega(\d\vec{Q}\w\vec{R}).
\ee
With the help of eq.\eq{quaddQ}, the equations of motion for $\vec{P}$ and $\vec{Q}$ of the action \eq{quadratic-tSDSen} can be worked out to be
\be\label{EOMP}
d\lrbrk{\hlf\bar{*}d\vec{P} + \vec{Q}} = 0,
\ee
\be\label{EOMQ}
\vec{R} = \hlf d\vec{P} - \hlf\bar{*}d\vec{P}.
\ee
The equation \eq{EOMP} implies
\be\label{EOMP-2}
\hlf\bar{*}d\vec{P} + \vec{Q}
=\hlf d\vec{\xi},
\ee
where $\vec{\xi}$ is a $2$-tuple $1$-form
which satisfies
\be\label{xiplusP}
d(\vec{\xi} + \vec{P}) = \bar{*}d(\vec{\xi} + \vec{P}).
\ee
Then from eq.\eq{EOMQ}-\eq{EOMP-2},
we obtain
\be\label{dHmdX}
d\vec{H}
=-d\vec{X}.
\ee
In spacetime with trivial topology,
this implies that $\vec{H}+\vec{X}$ is exact.
Note that from eq.\eq{vecH_defn} and eq.\eq{EOMQ}-\eq{EOMP-2},
the on-shell value for $\vec{H}$ is given by
$\vec{H} = d(\vec{\xi} - \vec{P})/2 - \vec{X}$.

From the analysis above, 
the system described by the action \eq{quadratic-tSDSen}
has uncoupled on-shell degrees of freedom which are $\vec{\xi} + \vec{P}$
and $\vec{\xi} - \vec{P}$.
The field $\vec{\xi} + \vec{P}$, according to eq.\eq{xiplusP}, has field strength which is twisted self-dual with respect to $\bar{*}$.
One may identify $\bar{A}^I_\m = (\xi^I_\m + P^I_\m)/2$.
As for the field $\vec{\xi} - \vec{P}$,
it satisfies
\be\label{nl_tsd_EOM}
*\lrbrk{\hlf d(\vec{\xi} - \vec{P}) - \vec{X}}
=\hlf d(\vec{\xi} - \vec{P}) - \vec{X},
\ee
i.e. its field strength is twisted self-dual with respect to $*$.
One may identify $A^I_\m = (\xi^I_\m - P^I_\m)/2.$
It can be seen using Hamiltonian analysis in subsection \ref{subsec:Ham}
that $\bar{A}^I_\m$ is unphysical
since its kinetic energy has the wrong sign,
whereas $A^I_\m$ is physical
with the correct sign of the kinetic energy.
It will also be seen that the unphysical and physical degrees of freedom
are uncoupled also at the Hamiltonian level
and that
the external fields $\bar{g}_{\m\n}, \bar{\t}$
couple only to the unphysical degrees of freedom,
whereas the external fields $g_{\m\n}, \t, \vec{X}$
couple only to the physical degrees of freedom.

\subsection{Non-linear twisted self-duality in the Sen formalism}\label{subsec:TwistedSDSen}
In order to generalise the action \eq{quadratic-tSDSen}
to describe a system with non-linear twisted self-duality,
one demands that the twisted self-duality condition \eq{quad_twisted_sd}
is generalised to a non-linear twisted self-duality condition
which should be self-consistent,
and that the equations of motion of the generalised action
should still imply eq.\eq{dHmdX}.
In this subsection, we will show how these requirements
lead to the required non-linear Sen action.

In the non-linear theory, the composite fields $\vec{R}, \vec{H}$
should now satisfy eq.\eq{vecR_defn}-\eq{vecR_asd},
as well as
\be\label{twisted_sd}
*\vec{H} = \vec{\cV}(\vec{H},\t,g_{\m\n}),
\ee
which generalises eq.\eq{quad_twisted_sd}.
The form of $\vec{\cV}$ should be such that
eq.\eq{twisted_sd} is self-consistent.
In other words, eq.\eq{twisted_sd}
should be equivalent to the consistutive relation
of the corresponding one-gauge-field theory.
This means that when expressing $\vec{H} = (F,G)$,
eq.\eq{twisted_sd} should take the form of
the constitutive relation
\be\label{consti_gen}
\tilde{G}_{\m\n} = \cA F_{\m\n} + \cB\tilde{F}_{\m\n},
\ee
where
\be
\tilde{G}_{\m\n} \equiv \hlf \sqrt{-g}\e_{\m\n\r\s}g^{\r\r'}g^{\s\s'}G_{\r'\s'},\qquad
\tilde{F}_{\m\n} \equiv \hlf \sqrt{-g}\e_{\m\n\r\s}g^{\r\r'}g^{\s\s'}F_{\r'\s'},
\ee
while $\cA, \cB$ are functions of $F_{\m\n}, g_{\m\n}, \t = C_0 + ie^{-\f}$
in such a way that $(F, G)$ satisfy
the generalised Gaillard-Zumino condition (c.f. eq.\eq{Gen-GZ})
\be\label{Gen-GZ_2}
(e^{-\f} + C_0^2 e^\f)F_{\m\n}F_{\r\s}\e^{\m\n\r\s}
-2C_0 e^\f F_{\m\n}G_{\r\s}\e^{\m\n\r\s}
+e^\f G_{\m\n}G_{\r\s}\e^{\m\n\r\s}
=0.
\ee

Another requirement for eq.\eq{twisted_sd} is that
both sides of the equation should transform in the same way
under $SL(2,\RR)$ duality transformation.
We will show in subsection \ref{subsec:Rnl} that
the composite field
$\vec{H}$ still transforms linearly under 
$SL(2,\RR)$ duality transformation.
That is, even in non-linear theories,
eq.\eq{SL2_1}-\eq{SL2_2}
still imply eq.\eq{RH_duality_trafo}.
There are two independent $SL(2,\RR)$ invariants
constructed out of $\vec{H}$ which are $[H^2]$ and $[H^4]$,
where
\be\label{H2_curved}
[H^2] \equiv M_{IJ}(H^2)^{IJ},\qquad
(H^2)^{IJ} \equiv H^{I}_{\r\s} H^J_{\n\m} g^{\m\r}g^{\n\s},
\ee
\be\label{H4_curved}
[H^4] \equiv M_{IJ}(H^4)^{IJ},\qquad
(H^4)^{IJ}\equiv (H^2)^{IK}M_{KL}(H^2)^{LJ}.
\ee
Furthermore
$*\vec{H}$
and
\be\label{H3_curved}
\vec{H}^3\equiv (H^2)^{IJ}M_{JK}H^K\vec{u}_I
\ee
also transform linearly under 
$SL(2,\RR)$ duality transformation,
i.e.
\be
(*\vec{H})'
=
(*\vec{H})
\begin{pmatrix}
A & C\\
B & D
\end{pmatrix},\qquad
(\vec{H}^3)'
=
\vec{H}^3
\begin{pmatrix}
A & C\\
B & D
\end{pmatrix}.
\ee
Let us therefore write eq.\eq{twisted_sd}
by using an ansatz
\be\label{ansatz_starH}
*\vec{H} = K_1\vec{H} + K_2\vec{H}^3,
\ee
where $K_1, K_2$ are functions of $[H^2]$ and $[H^4]$.
In fact, by applying $\Omega(*\vec{H}\w\cdot)$
on eq.\eq{ansatz_starH}, we see that
$[H^2]$ and $[H^4]$ are related by
\be
[H^2]K_1 + [H^4]K_2 = 0.
\ee
So we may say that $K_1$ and $K_2$ depend only on $[K^2]$.
Requiring that eq.\eq{consti_gen}
and eq.\eq{ansatz_starH}
are equivalent gives
\be\label{consti_vs_tsd_1}
\cA(K_1 e^{-\f} + K_2 C_0 F_{\m\n}G^{\m\n} - K_2 G^{\m\n}G_{\m\n}) -C_0(\cB - C_0)
+ e^{-2\f} = 0,
\ee
\be\label{consti_vs_tsd_2}
K_{2} \cA \left(-C_0 F_{\m\n}F^{\m\n} + F_{\m\n}G^{\m\n}\right)
+\cB-C_0
=0,
\ee
along with the Gaillard-Zumino condition eq.\eq{Gen-GZ_2}.
This means that there are three equations, eq.\eq{Gen-GZ_2}, \eq{consti_vs_tsd_1}-\eq{consti_vs_tsd_2}
relating four quantities $\cA, \cB, K_1, K_2$.
These equations can then be reduced
to one equation relating two quantities.
To do this, one
substitutes $\cA$ and $\cB$ obtained from
eq.\eq{consti_vs_tsd_1}-\eq{consti_vs_tsd_2}
into eq.\eq{Gen-GZ_2}.
This leads to
\be\label{1K1K2}
1 = \lrbrk{K_1 + \hlf[H^2]K_2}\lrbrk{K_1 + [H^2]K_2}.
\ee
The equation \eq{1K1K2} suggests that one may define
\be\label{Kdefn}
K \equiv K_1 + [H^2]K_2
=\lrbrk{K_1 + \hlf[H^2]K_2}^{-1}.
\ee
One may then express $K_1, K_2$ in terms of $K$
as follows
\be
K_1 = \frac{2}{K} - K,\qquad
K_2 = \frac{2}{[H^2]}\lrbrk{\frac{K^2 - 1}{K}}.
\ee

Therefore, consistent non-linear twisted self-duality condition
for $\vec{H}$ is of the form
\be\label{twisted_sd_2}
*\vec{H} = \vec{\cV} = \frac{2-K^2}{K}\vec{H} + \frac{2}{[H^2]}\lrbrk{\frac{K^2 - 1}{K}}\vec{H}^3,
\ee
where $K$ is a function of $[H^2]$.
Different forms of $K([H^2])$ would correspond to different non-linear theories.
The exception is when $K = 0$ which makes eq.\eq{twisted_sd_2} singular.
Furthermore, it may look as if eq.\eq{twisted_sd_2} would also be problematic if $[H^2] = 0$.
However, this singularity is in fact removable.
The theory with $[H^2] = 0$ has linear twisted self-duality condition $*\vec{H} = \vec{H}$ in which case, $K = 1$.
If a theory has twisted self-duality condition as a perturbation to the linear one, then $K = 1 + c_0[H^2] + \cdots$.
The twisted self-duality condition in this case reads $*\vec{H} = \vec{H} + c_0(-3[H^2]\vec{H} + 4\vec{H}^3) + {\cal O}(H^4).$
The singularity at $[H^2] = 0$ is clearly removed.

The equation \eq{twisted_sd_2} implies
\be\label{cVHU}
\Omega(\vec{\cV}\w\d_Q\vec{\cV})
-\Omega(\vec{H}\w\d_Q\vec{H})
=-4\sqrt{-g}\d_Q U d^4x,
\ee
where $U = U([H^2])$
is a function whose derivative with respect to $Q$ is given by
\be
\d_Q U
=-\ove{8}\frac{[H^2]}{K^2 - 1}K'\d_Q[H^2].
\ee
So derivative of $U$ with respect to $[H^2]$ is
\be\label{Uprime}
U'
=-\ove{8}\frac{[H^2]}{K^2 - 1}K'.
\ee
The equation \eq{cVHU}
is equivalent to
\be\label{nldQ}
\d_Q\lrbrk{\hlf\Omega\lrbrk{\vec{Q}\w\vec{R} - \vec{H}\w\vec{X}}
+d^4x\sqrt{-g}U([H^2])}
=\Omega(\d\vec{Q}\w\vec{R}),
\ee
which generalises eq.\eq{quaddQ}.
We have seen previously that
eq.\eq{quaddQ} allows us to realise that
the equations of motion of the action \eq{quadratic-tSDSen}
imply eq.\eq{dHmdX}.
In this non-linear case, we demand that the equations of motion
should still imply eq.\eq{dHmdX}.
Therefore,
the non-linear Sen action should take the form
\be\label{nl-tSDSen}
S_{\textrm{Sen}}
=-\int\Omega\lrbrk{\ove{4}d\vec{P}\w\bar{*}d\vec{P} - \vec{Q}\w d\vec{P} + \hlf\vec{Q}\w\vec{R} - \hlf\vec{H}\w\vec{X}}
-\int d^4x\sqrt{-g}U([H^2]).
\ee
Recall that $\vec{P}$ and $\vec{Q} = \bar{*}\vec{Q}$ are independent fields
whereas $\t, \bar{\t}, \vec{X}, g_{\m\n}, \bar{g}_{\m\n}$ are background fields.
The composite fields $\vec{R}, \vec{H}$ satisfy eq.\eq{vecR_defn}-\eq{vecR_asd}, \eq{twisted_sd_2}. The function $U([H^2])$ satisfies eq.\eq{Uprime}.

Unlike the linear case,
it is not a simple task
in non-linear case
to solve the equations \eq{Qsd}, \eq{vecH_defn}-\eq{vecR_asd}, \eq{twisted_sd_2}
to obtain the expression of $\vec{R}$ and $\vec{H}$
in terms of independent fields.
This in turn make
the writing of the action \eq{nl-tSDSen}
explicitly in terms of independent fields
even more complicated, if at all possible.
Nevertheless, when proving symmetries
and deriving equations of motion of the action \eq{nl-tSDSen},
the explicit expressions
in terms of independent fields
are not required.
So we postpone the discussion on
these expressions
to subsection \ref{subsec:RHQ}.

We will leave it to subsections \ref{subsec:Rnl}
and \ref{subsec:diffeo}
where we will show that
the action \eq{nl-tSDSen}
is invariant under 
$SL(2,\RR)$ duality transformation
and under diffeomorphism.

The analysis at the level of equations of motion
of the non-linear theory \eq{nl-tSDSen}
is the same as in the case of the quadratic theory \eq{quadratic-tSDSen}.
In particular, we still have eq.\eq{EOMP}-\eq{dHmdX} without modification.
The equation \eq{dHmdX} reflects
one of the standard features of the Sen formalism that
even in non-linear theory
$\vec{H}$ is exact on-shell, up to the external field $\vec{X}$, in spacetime with trivial topology. 
The conclusion similar to the case of quadratic action applies,
i.e. that the system is separated into physical and unphysical sectors
which are uncoupled from each other.
The field $(\vec{\xi} - \vec{P})/2$ is physical with correct sign of kinetic energy
and satisfies the modification of eq.\eq{nl_tsd_EOM}
which is now
\be\label{twisted_sd_os}
*\lrbrk{\hlf d(\vec{\xi} - \vec{P}) - \vec{X}}
=\vec{\cV}\lrbrk{\hlf d(\vec{\xi} - \vec{P}) - \vec{X}, \t, g_{\m\n}}.
\ee
That is, the field $(\vec{\xi} - \vec{P})/2$ satisfies non-linear twisted self-duality with respect to $g_{\m\n}, \t$.
The field $(\vec{\xi} + \vec{P})/2$ is unphysical with incorrect sign of kinetic energy and that it still satisfies linear twisted self-duality condition \eq{xiplusP} with respect to $\bar{g}_{\m\n}, \bar{\t}$.
In subsection \ref{subsec:Ham},
the justification at the Hamiltonian level will be shown.
In particular, the separation of the two sectors
and the verification of the sign of kinetic energy
are discussed.

Consider systems described by actions in the form of eq.\eq{nl-tSDSen},
i.e. 
those with $K = K([H^2]) \neq 0$ being specified.
Due to the separation of each system into physical and unphysical sectors,
we may focus only on the physical sector.
Then we ask for the corresponding one-gauge-field action
such that it also encodes the dynamics of this sector.

We are particularly interested in systems
whose twisted self-duality condition can be expressed as a perturbative expansion
of the linear twisted self-duality condition, with $K = 1, U = 0$.
We therefore consider theories where $K$ can be expanded as
\be\label{Kpert}
K = 1 + \frac{b_0}{4}[H^2] + \lrbrk{\frac{b_0^2}{32} + \frac{b_1}{4}}[H^2]^2 + {\cal O}([H^5]),
\ee
where $b_0,b_1\in\RR$.
The presentation of the form \eq{Kpert}
is simply for demonstration purpose.
Higher order terms can also be included up to any desirable order.
If higher order terms are considered, further free parameters
apart from $b_0$ and $b_1$ are introduced.
The set of free parameters characterises different non-linear theories.
In fact, one may also call the coefficients of $[H^2]$ and $[H^2]^2$ as $c_0$ and $c_1$.
However, our choice of coefficients in eq.\eq{Kpert} makes later expressions look simpler.
Next, having fixed $K$, we can use 
eq.\eq{twisted_sd_2} to determine twisted self-duality condition
and use eq.\eq{Uprime}
to determine $U$.
We obtain
\be\label{Vpert}
*\vec{H}
=\lrbrk{1 - \frac{3}{4}b_0[H^2] + \lrbrk{\frac{b_0^2}{32} - \frac{3b_1}{4}}[H^2]^2}\vec{H} + (b_0 + b_1[H^2])\vec{H}^3 + {\cal O}(H^6).
\ee
\be\label{Upert}
U = -\ove{16}[H^2] - \frac{b_1}{32b_0}[H^2]^2 + {\cal O}([H^2]^3).
\ee
The expression \eq{Upert} is valid for $b_0 \neq 0$.
If $b_0 = 0$, the form of $U$ would be different from the one given by eq.\eq{Upert}
and can easily be obtained by a perturbative analysis.
Next, by perturbatively solving for $\cA$ and $\cB$ in eq.\eq{consti_vs_tsd_1}-\eq{consti_vs_tsd_2},
substituting into eq.\eq{consti_gen}
and then solving
\be\label{bullG}
(\bullet G)^{\m\n}
=-\frac{2}{\sqrt{-g}}\frac{\d S}{\d F_{\m\n}} - (\bullet\hat{X})^{\m\n}
\ee
for the one-gauge-field action $S$
encoding the dynamics of the physical sector
we obtain, after renaming $F$ to $f$,
\be
S = \int d^4 x\cL,
\ee
where
\be\label{one_field_pert}
\begin{split}
\cL
&= 
\ove{8}C_0 f_{\m\n}\e^{\m\n\r\s}f_{\r\s}
+ \ove{4} f_{\m\n}\e^{\m\n\r\s}\hat{X}_{\r\s}\\
&\qquad+ \sqrt{-g}\lrbrk{y_1 + b_0 y_2 + 4b_0^2 y_1 y_2 + 4b_0y_2(4b_0^2 y_1^2 + b_0^2 y_2 - 2b_1 y_2)} + {\cal O}(f^9),
\end{split}
\ee
with $f_{\m\n} = \pa_\m A_\n - \pa_\n A_\m - X_{\m\n},$
and
\be
y_1 = -\frac{e^{-\f}}{4}f_{\m\n}f^{\m\n},\qquad
y_2 = \frac{1}{8}(e^{-\f}f_{\m\n}f^{\m\n})^2 + \frac{1}{8}(e^{-\f}f_{\m\n}\tilde{f}^{\m\n})^2.
\ee
When all the parameters $b_0, b_1, \cdots$
are set to zero, the action with Lagrangian \eq{one_field_pert}
reduces to eq.\eq{1field_Mtau},
as expected.

One may as well carry out the procedure above in reverse.
That is, given a one-gauge-field action
for a system with duality symmetry,
the non-linear duality symmetric Sen action
can be obtained.
An important example system
is the one whose one-gauge-field action
is the DBI action
\be\label{DBI}
S
=-\int d^4 x\sqrt{-\det(g_{\m\n} + e^{-\f/2}f_{\m\n})}
+\int (f\w \hat{X} + \hlf f\w f\w C_0).
\ee
By renaming $f$ to $F$ and considering eq.\eq{bullG},
one obtains
\be\label{consti_D3}
\tilde{G}
=C_0\tilde{F} - \frac{e^{-\f}F + \ove{4}e^{-2\f}[F\Ft]\Ft}{\sqrt{1 - \hlf e^{-\f}[F^2] -\ove{16}e^{-2\f}[F\tilde{F}]^2}}.
\ee
Comparing with eq.\eq{consti_gen},
the quantities $\cA, \cB$ in this case can be read off.
Then after substituting the obtained $\cA, \cB$
into eq.\eq{consti_vs_tsd_1}-\eq{consti_vs_tsd_2},
one obtains
\be
\begin{split}
K_1
&=\frac{1 - \hlf[F^2]e^{-\f} + \ove{8}[F^2]^2e^{-2\f} + \ove{16}[F\Ft]^2 e^{-2\f}}{\lrbrk{1 - \ove{4}[F^2]e^{-\f}}\sqrt{1 - \hlf e^{-\f}[F^2] -\ove{16}e^{-2\f}[F\tilde{F}]^2}}\\
&=\frac{1 - \frac{1}{4}[H^2]}{\sqrt{1 - \frac{1}{8}[H^2]}},
\end{split}
\ee
\be
\begin{split}
K_2
&= \frac{\sqrt{1 - \hlf e^{-\f}[F^2] -\ove{16}e^{-2\f}[F\tilde{F}]^2}}{4\lrbrk{1 - \ove{4}[F^2]e^{-\f}}}\\
&= \frac{1}{4\sqrt{1 - \frac{1}{8}[H^2]}}.
\end{split}
\ee
Then the twisted self-duality condition is
\be\label{tsd_DBI}
*\vec{H}
=\frac{1 - \frac{1}{4}[H^2]}{\sqrt{1 - \frac{1}{8}[H^2]}} \vec{H}
+\frac{1}{4\sqrt{1 - \frac{1}{8}[H^2]}}\vec{H}^3.
\ee
Note that the derivation leading to this condition essentially follows similar idea of \cite{Cederwall:1997ab}.
Then by using eq.\eq{Kdefn},
we obtain
\be\label{K_DBI}
K = \lrbrk{1-\frac{1}{8}[H^2]}^{-\hlf}.
\ee
Finally, using eq.\eq{Uprime}
gives rise to
\be\label{U_DBI}
U = \sqrt{1-\frac{1}{8}[H^2]}.
\ee

\subsection{The duality-symmetric D3-brane action in the Sen formalism}\label{subsec:SenD3}
In order to extend the action \eq{nl-tSDSen}
to describe D3-brane, one should
identify external fields with pull-backs of the background target space superfields
and specify the function $U$ and the corresponding twisted self-duality condition.
The fields $g_{\m\n}, C_0, \f$ are as given in subsection \ref{sec:standardD3}.
The external field $\vec{X}$ should be identified
with a pair of the pull-back of NS-NS super B-field
and the pull-back of $C_2$ superfield. That is,
\be
\vec{X}_2 = \vec{C}_2 = (B_2,C_2).
\ee
The function $U$ is given by eq.\eq{U_DBI}, whereas the twisted self-duality condition
is given by eq.\eq{tsd_DBI}.
Furthermore, one should also add
a term ${C_4 - (1/2)B_2\w C_2}$
to arrive at 
the duality-symmetric D3-brane action in the Sen formalism
\be\label{SenD3}
\begin{split}
S_{\textrm{Sen-D3}}
&=-\int\Omega\lrbrk{\ove{4}d\vec{P}\w\bar{*}d\vec{P} - \vec{Q}\w d\vec{P} + \hlf\vec{Q}\w\vec{R} - \hlf\vec{H}\w\vec{C}}
-\int d^4x\sqrt{-g}\sqrt{1-\frac{1}{8}[H^2]}\\
&\quad+\int \lrbrk{C_4 - \hlf B_2\w C_2}.
\end{split}
\ee
Recall that $\vec{P}$ and $\vec{Q} = \bar{*}\vec{Q}$
are independent fields, whereas $\vec{R}$ and $\vec{H}$ are composite fields
satisfying eq.\eq{vecR_defn}-\eq{vecR_asd}, \eq{twisted_sd}.
Note that the appearance of the $2$-tuple $2$-form source $\vec{X}_2 = \vec{C}_2$ within the expression of $\vec{H}$, i.e. $\vec{H} = \vec{H}|_{\vec{X} = \vec{0}} - \vec{X}$,
resembles the feature which is often appear in the duality-symmetric D3-brane actions, for example in \cite{Cederwall:1997ab}, which originally proposes this form, as well as in the PST formalism in \cite{Nurmagambetov:1998gp}, \cite{Suzuki:1999aa}, \cite{Vanichchapongjaroen:2017zuh}.

The action \eq{SenD3} is invariant under the transformation
\be\label{sg_1}
\d\vec{P} = -\vec{\l},\qquad
\d\vec{Q} = -\lrbrk{\frac{1+\bar{*}}{2}}d\d\vec{P},
\ee
\be
\d\vec{C}_2 = d\vec{\l},\qquad
\d C_4 = d\m_3 + B_2\w d\m_1,
\ee
\be\label{sg_3}
\d M_{IJ} = \d g_{\m\n} = \d\bar{M}_{IJ} = \d\bar{g}_{\m\n}
=0.
\ee

The action \eq{SenD3} also has kappa-symmetry.
Since the theory couples to type IIB supergravity background,
the kappa-symmetry variation of the pull-back of the background superfields
satisfy eq.\eq{ikE-1}-\eq{ikR5}. Then let us suppose that the kappa-symmetry variations
for unphysical fields vanish.
This means that
\be\label{dkQ}
\d_\k\vec{Q} = -\lrbrk{\frac{1+\bar{*}}{2}}d\d_\k\vec{P},
\ee
\be
\d_\k \bar{g}_{\m\n} = 0 = \d_\k \bar{M}_{IJ}.
\ee
So far, the variation $\d_\k S_{\textrm{Sen-D3}}$ contains the expression $i_\k\vec{C}$
which is unwanted.
By setting
\be\label{dkP}
\d_\k\vec{P} = -i_\k\vec{C},
\ee
the expressions containing $i_\k\vec{C}$ within $\d_\k S_{\textrm{Sen-D3}}$
are cancelled out. We obtain
\be\label{dkSSenD3}
\d_\k S_{\textrm{Sen-D3}}
=\int d^4 x\ i\sqrt{-g}\bar{E}_\m W_1^\m(1-\bar{\G})i_\k E
+\int d^4 x \ove{8}\sqrt{-g}\bar{\L}W_2 (1-\bar{\G})i_\k E,
\ee
where
\be
W_1^\m
=\bar{\g}\g^\m\t_2
+\Omega_{IJ}\tilde{H}^{J\m\n}\g_\n T^I_-
+\hlf\Omega_{IJ}\tilde{H}^{I\r(\m} H^{|J|}_{\r\s}g^{\n)\s}
-\sqrt{1-\frac{1}{8}[H^2]}\g^\m,
\ee
\be
W_2
=2\Omega_{IJ}\tilde{H}^{J\m\n}\g_{\m\n}T^I_+
-\Omega_{MI}H^{I}_{\m\n}\tilde{H}^{J\m\n}M_{JK}T^K_+ T^M_-,
\ee
\be\label{barG_SenD3}
\bar{\G}_{\textrm{Sen-D3}} = \frac{-\bar{\g}\t_2 +\ove{4}\Omega_{IJ}\tilde{H}^{J\m\n}\g_{\m\n}T^I_-}{\sqrt{1-\frac{1}{8}[H^2]}},
\ee
with
\be
\tilde{H}^{J\m\n}
=\hlf\frac{\e^{\m\n\r\s}}{\sqrt{-g}}H^{J}_{\r\s},
\ee
\be
\vec{T}_{\pm} = (e^{\f/2}\t_3, \pm e^{-\f/2}\t_1 + e^{\f/2}C_0\t_3).
\ee
Spacetime indices are lowered and raised by the metric $g_{\m\n}$ and its inverse $g^{\m\n}$.
The matrix $\bar{\G}_{\textrm{Sen-D3}}$ satisfies $(\bar{\G}_{\textrm{Sen-D3}})^2 = 1, \tr(\bar{\G}_{\textrm{Sen-D3}}) = 0$.
In order for the variation \eq{dkSSenD3}
to vanish, one should set
\be
i_\k E^{\ua} = ((1 + \bar{\G}_\textrm{Sen-D3})\k)^{\ua}.
\ee
Half of the components of $\k$
can be used to gauge away some of the fermionic degrees of freedom
and to realise the matching between bosonic and fermionic degrees of freedom.
This is in the way similarly to 
the one discussed at the end of subsection \ref{sec:standardD3}.

\section{Properties}\label{sec:properties}

\subsection{Duality symmetry of the non-linear action}\label{subsec:Rnl}
As analysed in subsection \ref{subsec:quad_TwistedSDSen},
the proof that the quadratic action \eq{quadratic-tSDSen} is duality symmetric
proceeds as follows.
We first express
$\vec{R}$ and $\vec{H}$
in terms of other fields,
work out their transformation under duality transformation
and finally show that the action itself is duality invariant.
However, if we generalise to the non-linear theory
with action \eq{nl-tSDSen},
the same approach would be too involved.
This is because
it is not simple to express
$\vec{R}$ and $\vec{H}$
in terms of independent fields
(these expressions are to be discussed later
in subsection \ref{subsec:RHQ}).
A new strategy is therefore required.

In this subsection,
we will consider a class of non-linear theories \eq{nl-tSDSen}
which are perturbations to the theory described by the quadratic
action \eq{quadratic-tSDSen}.
The proof that the actions are duality invariant
can be done
by first
considering the equations \eq{Qsd}, \eq{vecH_defn}-\eq{vecR_asd}, \eq{twisted_sd_2}
as well as the perturbative expansion of $\vec{R}$ and $\vec{H}$
as polynomials in $\vec{Q}$.
In particular, the equation \eq{twisted_sd_2}
can be perturbatively expanded,
giving equations characterised by the order of $\vec{Q}$.
These equations can be used to directly show that
each perturbation order of $\vec{R}$ and $\vec{H}$
transform
in the standard way
under the duality transformation.
This result can then be used to easily show that
the actions are duality invariant.
The class under consideration contains
the theory which describes a DBI system
which can in turn be extended to describe a D3-brane coupled to type IIB supergravity background.

Consider transformations of fields under
$SL(2,\RR)$ duality transformation
with parameters $A, B, C, D\in\RR$ with $AD - BC = 1$.
The fields $\t = C_0 + ie^{-\f}$ and $\bar{\t} = \bar{C}_0 + ie^{-\bar{\f}}$
transform under the duality transformation as in eq.\eq{SL2_2}.
Let $\vec{L}$ be any $2$-tuple $2$-form.
Furthermore, let $\vec{L}$ transform under the duality transformation
as
\be
L'^I = L^J S_J{}^I,
\ee
where
\be
S_J{}^I
\equiv
\begin{pmatrix}
A & C\\
B & D
\end{pmatrix}.
\ee
With this property, we state that 
$\vec{L}$ is linear under $SL(2,\RR)$ duality transformation.
Then with $M_{IJ}$ as defined by eq.\eq{M_mat},
it can be seen that $L_I\equiv M_{IJ}L^J$ transforms as
\be
L'_I = (S^{-1})_I{}^JL_J.
\ee
Now let
$\vec{L}_1, \vec{L}_2$
be any $2$-tuple $2$-forms
which are also linear under $SL(2,\RR)$ duality transformation.
Then
\be
[L'L'_{1}]
\equiv L'^I_{\m\n}L'^{J\n\m}M'_{IJ}
=L^I_{\m\n}L^{J\n\m}M_{IJ}
\equiv [LL_1],
\ee
i.e. $[LL_1]$ is invariant under $SL(2,\RR)$ duality transformation.
It can also be shown that
$[L^I L_1^J]M_{JK}L_2^K\vec{u}_I \equiv L^I_{\m\n} L_1^{J\n\m}M_{JK}L_2^K\vec{u}_I$ is linear under $SL(2,\RR)$ duality transformation.

Consider twisted self-duality condition \eq{twisted_sd_2}.
By expanding $K = 1 + \sum_{n = 0}^\infty c_n [H^2]^{n+1}$,
we obtain
\be\label{tsd_expand}
*\vec{H}
=\lrbrk{1+\sum_{n=0}^\infty a_n[H^2]^{n+1}}\vec{H} + \sum_{n=0}^\infty b_n[H^2]^{n}\vec{H}^3,
\ee
where the coefficients $\{a_n\}, \{b_n\}$ can be given in terms of $\{c_n\}$.
For our purpose in this subsection, it is not necessary to
use the explicit relationships of these coefficients. 
It is convenient to use $\vec{Q}_X, \vec{R}_X$ as defined in eq.\eq{QXRX}.
Let us expand
\be\label{Hexpand}
\vec{R}_X = \sum_{k = 1}^\infty \vec{R}_{X(k)},\qquad
\vec{H} = \sum_{k = 1}^\infty \vec{H}_{(k)},
\ee
where $\vec{R}_{X(k)}, \vec{H}_{(k)}$ are polynomials of order $k$ in $\vec{Q}_X$.
Substituting eq.\eq{Hexpand} into eq.\eq{vecH_defn}
gives
\be\label{HkQR}
\vec{H}_{(k)}
=\d_{k1}\vec{Q}_X - \vec{R}_{X(k)},
\ee
for $k = 1,2,3,\cdots$.
By substituting eq.\eq{Hexpand}
into eq.\eq{tsd_expand} and focusing on the second order term,
we see that
\be
*\vec{H}_{(2)} = \vec{H}_{(2)}.
\ee
Then by using eq.\eq{HkQR} and eq.\eq{QXRX_tsd},
we obtain
\be
(1+*\bar{*})\vec{H}_{(2)} = \vec{0},
\ee
which implies $\vec{H}_{(2)} = \vec{0}$.
It then follows that $\vec{H}_{(2n)} = \vec{0}$ for $n = 1,2,3,\cdots$.
So
\be\label{Hexpand_2}
\vec{H} = \sum_{n = 0}^\infty \vec{H}_{(2n+1)}.
\ee
After substituting eq.\eq{Hexpand_2} into eq.\eq{tsd_expand},
the first order term is
\be\label{tsd1}
*\vec{H}_{(1)} = \vec{H}_{(1)}.
\ee
So by following a similar consideration
as in subsection \ref{subsec:quad_TwistedSDSen}
(in particular see eq.\eq{quad_twisted_sd}-\eq{HandHhat}, \eq{RH_duality_trafo}),
we can conclude that $\vec{H}_{(1)}$ is linear under $SL(2,\RR)$ duality transformation.
Next, consider the third order terms of eq.\eq{tsd_expand}.
These are (there is no $a_0$ on RHS since $[H_{(1)}H_{(1)}] = 0$)
\be\label{tsd3}
-(1-*)\vec{H}_{(3)}
=
b_0[H_{(1)}^I H_{(1)}^J]M_{JK}H_{(1)}^K\vec{u}_I.
\ee
Based on the result of the previous paragraph,
it can be seen that RHS of eq.\eq{tsd3} is 
linear under $SL(2,\RR)$ duality transformation.
So $-(1-*)\vec{H}_{(3)}$ is also linear under $SL(2,\RR)$ duality transformation.
By
expressing $\vec{H}_{(k)} = (F_{(k)},\hat{F}_{(k)})$, we have
\be\label{starHasF}
-(1-*)\vec{H}_{(3)}
=e^{\f}(\bullet\hat{F}_{(3)} - \bullet T^* F_{(3)})\vec{u}_1
+e^{\f}\bullet T(\hat{F}_{(3)} - T^*F_{(3)})\vec{u}_2,
\ee
where $T$ and $T^*$ are as defined in eq.\eq{Tdefn}-\eq{Tsdefn}.
Since $-(1-*)\vec{H}_{(3)}$ is linear under $SL(2,\RR)$ duality transformation, this means that
\be
e^{\f'}(\bullet\hat{F}'_{(3)} - \bullet T'^* F'_{(3)})
=Ae^{\f}(\bullet\hat{F}_{(3)} - \bullet T^* F_{(3)})
+Be^{\f}\bullet T(\hat{F}_{(3)} - T^*F_{(3)}),
\ee
and
\be
e^{\f'}\bullet T'(\hat{F}'_{(3)} - T'^*F'_{(3)})
=Ce^{\f}(\bullet\hat{F}_{(3)} - \bullet T^* F_{(3)})
+De^{\f}\bullet T(\hat{F}_{(3)} - T^*F_{(3)}).
\ee
These give
\be
(BT^* + A)\bullet \hat{F}'_{(3)} - \bullet(DT^* + C)F'_{(3)}
=\bullet\hat{F}_{(3)} - \bullet T^* F_{(3)}.
\ee
By comparing the coefficients of $C_0$ and $e^{-\f}$,
we can see that $\vec{H}_{(3)}$ is linear under $SL(2,\RR)$
duality transformation.

In general, by substituting eq.\eq{Hexpand_2} into eq.\eq{tsd_expand},
the $(2n+1)$th order terms can be put in the form
\be\label{H2n1}
-(1-*)\vec{H}_{(2n+1)} = \vec{W}_{(2n+1)}(\vec{H}_{(1)},\vec{H}_{(3)}\cdots,\vec{H}_{(2n-1)}).
\ee
The functions $\vec{W}_{(2n+1)}$ is of the form that if
$\vec{H}_{(1)},\vec{H}_{(3)}\cdots,\vec{H}_{(2n-1)}$ are all 
linear under $SL(2,\RR)$
duality transformation,
then $\vec{W}_{(2n+1)}(\vec{H}_{(1)},\vec{H}_{(3)}\cdots,\vec{H}_{(2n-1)})$
is also linear
under $SL(2,\RR)$
duality transformation.
Then by using the similar analysis as in the previous paragraph,
it can be concluded that 
$\vec{H}_{(2n+1)}$ is also 
linear under $SL(2,\RR)$
duality transformation.
We may then use mathematical induction
to conclude that
$\vec{H}$ at any perturbation order
is linear under $SL(2,\RR)$
duality transformation.
Then since $\vec{R}_X = \vec{Q}_X - \vec{H}$,
it is easy to see that $\vec{R}_X$
is also linear under $SL(2,\RR)$
duality transformation.
This finally implies that
the action \eq{nl-tSDSen}
is invariant under $SL(2,\RR)$
duality transformation.

\subsection{Diffeomorphism invariance}\label{subsec:diffeo}
The Sen chiral form actions
have been shown to have three types of related diffeomorphism symmetries \cite{Hull:2023dgp}
(see also \cite{Sen:2015nph}, \cite{Andriolo:2020ykk} for the original derivation of $\zeta$-transformation,
and \cite{Vanichchapongjaroen:2020wza} for extension of $\zeta$-transformation to non-linear cases).
The first type is the standard diffeomorphism, or $\xi$-transformation in which every fields transform under Lie derivatives $\d_\xi^{(standard)} = -\pounds_\xi$.
The second type is the $\zeta$-transformation $\d_\zeta^{(zeta)}$, in which the variations on $\bar{g}$-sector vanishes whereas the variations on $g$-sector
takes a form which reduces to $-\pounds_\zeta$ on-shell.
The third type is the $\chi$-transformation $\d_\chi^{(chi)}$,
which can be obtained by requiring that
the sum of the $\zeta$-transformation and the $\chi$-transformation 
with identified parameters give the standard diffeomorphism,
i.e. $\d_\chi^{(chi)} = \d_\chi^{(standard)} - \d_\chi^{(zeta)}$.

In this subsection, we are going to show that
the Sen duality-symmetric action \eq{nl-tSDSen} also has these three types
of diffeomorphism symmetries.
Similar to the case of Sen chiral form action,
the only non-trivial check is the $\zeta$-transformation
since
it is straightforward to see that the Sen duality-symmetric action
has the standard diffeomorphism symmetry
and that the $\chi$-transformation can be obtained from
$\d_\chi^{(chi)} = \d_\chi^{(standard)} - \d_\chi^{(zeta)}$.

The action \eq{nl-tSDSen} is invariant under the $\zeta$-type diffeomorphism $x^\m\mapsto x^\m + \zeta^\m$:
\be\label{dzeta_P}
\d^{(zeta)}_\zeta\vec{P} = i_\zeta(\vec{H}+\vec{X}),
\ee
\be\label{dQdP}
\d^{(zeta)}_\zeta\vec{Q} = -\hlf(1+\bar{*})d\d^{(zeta)}_\zeta\vec{P},
\ee
\be\label{dzeta_g}
\d^{(zeta)}_\zeta g_{\m\n} = -\pounds_\zeta g_{\m\n},\qquad
\d^{(zeta)}_\zeta M_{IJ} = -\pounds_\zeta M_{IJ},\qquad
\d^{(zeta)}_\zeta\vec{X} = -\pounds_\zeta\vec{X},
\ee
\be\label{dzeta_gbar}
\d^{(zeta)}_\zeta \bar{g}_{\m\n} = 0 = \d^{(zeta)}_\zeta\bar{M}_{IJ}.
\ee
In order to see this, we first consider a generic variation $\d$.
Direct calculation gives the following identities:
\be
\Omega(*\vec{H}\w\d\vec{H})
=-\ove{4}d^4x\sqrt{-g}\lrbrk{\d
-\d_{g,M}}[H^2],
\ee
\be
\Omega(*\vec{H}\w\d\vec{H}^3)
=-\ove{8}d^4x\sqrt{-g}
\lrbrk{3\d-\d_{g,M}}[H^4],
\ee
\be
\Omega(*\vec{H}^3\w\d\vec{H})
=-\ove{8}d^4x\sqrt{-g}
\lrbrk{\d-\d_{g,M}}[H^4],
\ee
where
\be
\d_{g,M}\equiv
\d g^{\m\n}\frac{\d}{\d g^{\m\n}}\bigg|_{\textrm{fix }M_{IJ}, H^I_{\m\n}}
+ \d M_{IJ}\frac{\d}{\d M_{IJ}}\bigg|_{\textrm{fix }g_{\m\n}, H^I_{\m\n}}.
\ee
Hence
\be\label{VVmHH-pre}
\Omega(\vec{\cV}\w\d\vec{\cV}) - \Omega(\vec{H}\w\d\vec{H})
=-4d^4 x \sqrt{-g}\d U ,
\ee
\be\label{VVpHH-pre}
\Omega(\vec{\cV}\w\d\vec{\cV}) + \Omega(\vec{H}\w\d\vec{H})
=d^4 x\sqrt{-g}\lrbrk{\hlf K_1\d_{g,M}[H^2] + \ove{4} K_2\d_{g,M}[H^4]}.
\ee
Applying $\d^{(zeta)}_\zeta$ on the action \eq{nl-tSDSen}
and using eq.\eq{dQdP}-\eq{dzeta_gbar}
gives
\be\label{deltaS_2}
\begin{split}
\d^{(zeta)}_\zeta S_{\textrm{Sen}}
&=\int\bigg(-\Omega(d\d^{(zeta)}_\zeta\vec{P}\w(\vec{H}+\vec{X})) + \hlf\Omega(\vec{X}\w\pounds_\zeta\vec{X}) - \Omega((\vec{H}+\vec{X})\w\pounds_\zeta\vec{X})\\
&\qquad+d^4 x\sqrt{-g}\lrbrk{-\pounds_\zeta U + \ove{8}K_1\d_{g,M}[H^2] + \ove{16}K_2\d_{g,M}[H^4]}
\bigg).
\end{split}
\ee
By using eq.\eq{VVmHH-pre}-\eq{VVpHH-pre} with $\d = -\pounds_\zeta$,
and applying them to eq.\eq{deltaS_2},
we obtain
\be
\begin{split}
\d^{(zeta)}_\zeta S_{\textrm{Sen}}
&=-\int\Omega(d(\d^{(zeta)}_\zeta\vec{P}-i_\zeta(\vec{H}+\vec{X}))\w(\vec{H}+\vec{X})),
\end{split}
\ee
which vanishes if we finally impose eq.\eq{dzeta_P}.

\subsection{Hamiltonian analysis}\label{subsec:Ham}
Let the spatial indices be labelled by lower-case Roman alphabets
for example $a,b,c,i,j,k\in\{1,2,3\}$.
The ADM decomposition is applied on the metrics $g_{\m\n}, \bar{g}_{\m\n}$
as
\be
g^{00} = -\ove{N^2},\qquad
g^{0i} = \frac{N^i}{N^2},\qquad
g^{ij} = \g^{ij} - \frac{N^i N^j}{N^2},
\ee
\be
\bar{g}^{00} = -\ove{\bar{N}^2},\qquad
\bar{g}^{0i} = \frac{\bar{N}^i}{\bar{N}^2},\qquad
\bar{g}^{ij} = \bar{\g}^{ij} - \frac{\bar{N}^i \bar{N}^j}{\bar{N}^2}.
\ee

In the index notation, the Lagrangian from eq.\eq{nl-tSDSen}
reads
\be
\begin{split}
\cL
=&\hlf\sqrt{-\bar{g}}\bar{M}_{IJ}\pa_\m P_\n^I\pa_\r P_\s^J\bar{g}^{\m[\r}\bar{g}^{\s]\n}
+\sqrt{-\bar{g}}\bar{M}_{IJ}Q^I_{\m\n}\pa_\r P_\s^J \bar{g}^{\m\r}\bar{g}^{\n\s}\\
&-\ove{4}\sqrt{-\bar{g}}\bar{M}_{IJ}Q^I_{\m\n} R_{\r\s}^J \bar{g}^{\m\r}\bar{g}^{\n\s}
+\ove{8}\Omega_{IJ}H^I_{\m\n} X_{\r\s}^J \e^{\m\n\r\s}
\\
&-\sqrt{-g}U(M_{IJ}H^I_{\m\n}H^J_{\s\r}g^{\m\r}g^{\n\s}).
\end{split}
\ee
The phase space variables are $P_\m^I, Q^I_{ij}$ and their conjugate momenta.
The scalar fields, the spacetime metrics, as well as $\vec{X}$ are all external fields.
One may use the twisted self-duality conditions (i.e. $\vec{Q} = \bar{*}\vec{Q}, \vec{R} = -\bar{*}\vec{R}$ and eq.\eq{twisted_sd})
to realise that
$Q^I_{\m\n}, R^I_{\m\n}, H^I_{\m\n}$ depend on $Q^I_{ij}$ and external fields.

Conjugate momenta for $P_\m^I$ are
\be
\p_I^0\equiv\frac{\pa\cL}{\pa\dot{P}_0^I} = 0,
\ee
\be
\begin{split}
\p_I^i
\equiv&\frac{\pa\cL}{\pa\dot{P}_i^I}\\
=&\sqrt{-\bar{g}}\bar{M}_{IJ}\pa_\r P_\s^J\bar{g}^{0[\r}\bar{g}^{\s]i}
+\sqrt{-\bar{g}}\bar{M}_{IJ}Q_{\m\n}^J\bar{g}^{\m 0}\bar{g}^{i\n}.
\end{split}
\ee
Conjugate momenta for $Q_{ij}^I$ are
\be
\Pi^{ij}_I
\equiv
\frac{\pa\cL}{\pa \dot{Q}^I_{ij}}
=0.
\ee
Therefore, the primary constraints are $\p^0_I\approx 0$, and $\Pi^{ij}_I\approx 0$.

The canonical Hamiltonian is
\be
\begin{split}
\cH
=&\p_I^i\pa_0 P_i^I-\cL + \zeta^I\p_I^0 + \th^I_{ij}\Pi^{ij}_I\\
=&-\frac{\bar{N}}{\sqrt{\bar{\g}}}\bar{\g}_{ij}\bar{M}^{IJ}\p^{i}_I\p^{j}_J
+2\p^{b}_I\bar{N}^a\pa_{[a} P^I_{b]}\\
&\qquad
-2\frac{\bar{N}}{\sqrt{\bar{\g}}}\bar{\g}_{ij}\p^{+i}_I\e^{jab}\bar{J}^I{}_J Q^J_{ab}
-\ove{2}\bar{N}\sqrt{\bar{\g}}\bar{M}_{IJ}\pa_a P_b^I\bar{\g}^{a[m}\bar{\g}^{n]b}\pa_m P^J_n\\
&\qquad
-\ove{2}\bar{N}\sqrt{\bar{\g}}\bar{M}_{IJ}Q^I_{ab}\bar{\g}^{a[m}\bar{\g}^{n]b}(Q-R)^J_{mn}\\
&\qquad
-\ove{4}\Omega_{IJ} H^I_{ij}X^J_{0k}\e^{ijk}
-\ove{4}\Omega_{IJ} H^I_{0k}X^J_{ij}\e^{ijk}\\
&\qquad+ \sqrt{-g}U
+ \p^i_I\pa_I P_0^I
+ \zeta^I\p_I^0 + \th^I_{ij}\Pi^{ij}_I,
\end{split}
\ee
where
\be
\p^{Ii}\equiv\bar{M}^{IJ}\p_J^i,\quad
\p^{\pm i}_I
\equiv\p_I^i\pm\hlf\Omega_{IJ}\e^{iab}\pa_a P^J_{b}.
\ee
Requiring the time derivative of the primary constraints
to vanish on the constraint surface
gives secondary constraints
\be\label{secondary_constraints}
\pa_i\p^i_I\approx 0,\qquad
H^I_{jk}-\e_{ijk}\Omega^{IJ}\p^{Xi}_J\approx 0,
\ee
where
\be
\p^{Xi}_I\equiv
\p^{+i}_I + \hlf\Omega_{IJ}X^J_{jk}\e^{ijk}.
\ee
Requiring the time derivative of the secondary constraints
to vanish on the constraint surface
does not produce further constraints.
After reclassifying the constraints, it turns out that
the first-class constraints are $\p_I^0\approx 0, \pa_i\p_I^i\approx 0$,
whereas the second-class constraints are 
$\Pi^{ij}_I\approx 0, H^I_{jk}-\e_{ijk}\Omega^{IJ}\p^{Xi}_J\approx 0$.

One may then reduce the phase space by eliminating
$\Pi^{ij}_I$ and $Q_{ij}^I$ by solving the second-class constraints.
After the elimination, the Hamlitonian becomes
\be\label{cH-reduced}
\begin{split}
\cH
=&-\hlf\frac{\bar{N}}{\sqrt{\bar{\g}}}\bar{\g}_{ij}\bar{M}^{IJ}\p^{-i}_I\p^{-j}_J
+\hlf\Omega_{IJ}\p^{-Ii}\p^{-Jj}\bar{N}^k\e_{ijk}\\
&-\hlf\p_I^{Xi} H_{0i}^I
-\p_I^{Xi}X_{0i}^I
+\ove{4}\Omega_{IJ}X_{0i}^I X_{jk}^J\e^{ijk}
+ \sqrt{-g}U
+ \p^i_I\pa_I P_0^I + \zeta^I\p_I^0,
\end{split}
\ee
where
it is understood that $H^I_{ij}$ within $U$
are to be replaced, due to the second weak equality of eq.\eq{secondary_constraints}, by $\e_{ijk}\Omega^{IJ}\p^{Xk}_J$,
whereas $H_{0i}^I$ are to be given through twisted self-duality conditions in terms of 
$H_{ij}^I$, which are in turn expressed in terms of $\p^{Xk}_J$.

There are then only the first-class constraints left.
A proper treatment is to impose gauge fixing conditions
so that all the first-class constraints become second-class ones,
which are then solved to eliminate some phase space variables.
We choose the gauge-fixing conditions
\be\label{g-fix}
P^I_0\approx 0,\qquad
\pa_i P^I_i\approx 0.
\ee
Note that the Coulomb conditions $\pa_i P^I_i\approx 0$ are not covariant
and also does not contain any metric.
The time derivatives of the conditions \eq{g-fix}
do not give rise to further constraints.
The reclassification of the constraints
suggest that the remaining constraints,
i.e. $\p_I^0\approx 0, \pa_i\p_I^i\approx 0, P^I_0\approx 0, \pa_i P^I_i\approx 0$
are all second-class.
These constraints are solved to eliminate
$\p_I^0, P^I_0$ and the longitudinal components
of $\p^i_I$ and $P_i^I$ (for each $I$).
Since $\p^i_I$ only have transverse components remaining,
by Helmholtz decomposition in spacetime with trivial topology,
there exist $\psi_{Ik}$ such that $\p^i_I = \e^{ijk}\pa_j\psi_{Ik}/2$.
The longitudinal components of $P_i^I$
are not present in the Hamiltonian \eq{cH-reduced}
as it depends on $P_i^I$ only through $\e^{ijk}\pa_j P_k^I$ which contribute to the transverse components.

Finally, the reduced Hamiltonian, with all constraints eliminated, is
\be
\begin{split}
\cH
=\cH_- + \cH_+,
\end{split}
\ee
where
\be
\begin{split}
\cH_-=&-\hlf\frac{\bar{N}}{\sqrt{\bar{\g}}}\bar{\g}_{ij}\bar{M}^{IJ}\p^{-i}_I\p^{-j}_J
+\hlf\Omega_{IJ}\p^{-Ii}\bar{N}^a\e_{aij}\p^{-Jj},
\end{split}
\ee
\be
\begin{split}
\cH_+=-\hlf\p_I^{Xi} H_{0i}^I
-\p_I^{Xi}X_{0i}^I
+\ove{4}\Omega_{IJ}X_{0i}^I X_{jk}^J\e^{ijk} + \sqrt{-g}U,
\end{split}
\ee
where $H_{\m\n}^I$ are treated as explained earlier
and that $\p^i_I = \e^{ijk}\pa_j\psi_{Ik}/2$.
The Hamiltonian $\cH$ is separated into two parts $\cH_-, \cH_+$.
The sign of $\cH_-$ is incorrect
which suggests that
$\psi_{Ii}-\Omega_{IJ} P^J_i$ is unphysical.
On the other hand,
$\cH_+$ has the correct sign,
which suggests that $\psi_{Ii}+\Omega_{IJ} P^J_i$ is physical.
The unphysical part $\cH_-$ contains $\psi_{Ii}-\Omega_{IJ} P^J_i, \bar{g}_{\m\n}, \bar{\t}$,
whereas the physical part $\cH_+$ contains $\psi_{Ii}+\Omega_{IJ} P^J_i, X^I_{ij}, g_{\m\n}, \t$.

The analysis of this subsection which shows the separation of the system into two sectors
at the Hamiltonian level is in agreement with the similar analysis at the equations of motion level given in subsections \ref{subsec:quad_TwistedSDSen}-\ref{subsec:TwistedSDSen}.

In the case of quadratic theory \eq{quadratic-tSDSen} whose twisted self-duality  condition is linear, i.e. $\vec{H}=*\vec{H}$,
the physical part of the Hamiltonian is
\be
\cH_+
=\hlf\frac{N}{\sqrt{\g}}\g_{ij}M^{IJ}\p^{Xi}_I\p^{Xj}_J
-\hlf\Omega_{IJ}\p^{XIi}N^k\e_{kij}\p^{XJj},
\ee
whose kinetic term clearly has positive sign.

\subsection{The form of $\vec{R}$ and $\vec{H}$ in terms of independent fields in non-linear theory}\label{subsec:RHQ}
As presented so far in this paper,
there is no need to express the form of 
the composite fields $\vec{R}$ and $\vec{H}$ in terms of independent fields
when constructing the non-linear $SL(2,\RR)$ duality-symmetric Sen action
and an $SL(2,\RR)$ duality-symmetric D3-brane Sen action,
as well as when verifying the symmetries of the actions,
deriving equations of motion,
and analysing constraints.
Nevertheless, the expression could be useful for some other purposes.
So in this subsection, we present an algorithm
to perturbatively obtain $\vec{R}$ and $\vec{H}$
in terms of the independent fields.
In general, the expressions are quite complicated.
So it is still unclear how to write the non-linear action
entirely in terms of independent fields.
However, in the case of flat metrics,
the expression of $\vec{R}$ and $\vec{H}$
are much simpler which in turn
make it simple to write
the non-linear action in terms of independent fields.

Let us start from the expanded form, eq.\eq{tsd_expand}, of twisted-self-duality condition of $\vec{H}$.
Recall that the coefficients $\{a_n\}, \{b_n\}$ are related.
For example, $a_0 = -3b_0/4$.
Next, we perturbatively expand $\vec{H}$ as in eq.\eq{Hexpand}-\eq{HkQR}.
In fact, as seen from eq.\eq{Hexpand_2},
$\vec{H}$ only contains odd orders.
We have already seen that
after substituting eq.\eq{Hexpand_2} into
eq.\eq{tsd_expand},
the first order expansion
is given by eq.\eq{tsd1},
which is simply the linear twisted self-duality condition.
So we may read off from eq.\eq{HandHhat}
to obtain
\be\label{F1}
F_{(1)} = (\bar{T} - T^*)^{-1}(\bar{T} - \bar{T}^*)Q_X,\qquad
\hat{F}_{(1)} = T^*(\bar{T} - T^*)^{-1}(\bar{T} - \bar{T}^*)Q_X,
\ee
where
recall that $\vec{Q}_X$ is defined in eq.\eq{QXRX}
and we define
$F_{(1)}$ and $\hat{F}_{(1)}$ through $\vec{H}_{(1)} = (F_{(1)},\hat{F}_{(1)})$.
More generally, we write $\vec{H}_{(k)} = (F_{(k)},\hat{F}_{(k)})$.
Consider the third order of eq.\eq{tsd_expand} which is given by eq.\eq{tsd3}.
By using eq.\eq{starHasF},
we obtain
\be\label{tsd3_int}
e^{\f}(\vec{u}_1 + \vec{u}_2T)(\bullet\hat{F}_{(3)} - \bullet T^* F_{(3)})
=
b_0[H_{(1)}^I H_{(1)}^J]M_{JK}H_{(1)}^K\vec{u}_I.
\ee
Note from eq.\eq{HkQR} that $\vec{H}_{(2n+1)} = -\vec{R}_{X(2n+1)}$ for $n>0$.
Then $\vec{H}_{(2n+1)} = -\bar{*}\vec{H}_{(2n+1)}$ for $n>0$.
This implies that $\hat{F}_{(2n+1)} = \bar{T}F_{(2n+1)}$ for $n>0$.
This simplifies LHS of eq.\eq{tsd3_int}.
Then along with direct calculation of RHS of eq.\eq{tsd3_int},
we obtain
\be\label{tsd3_int2}
e^{\f}(\vec{u}_1 + \vec{u}_2T)\bullet(\bar{T}-T^*)F_{(3)}
=
b_0e^{-\f}(\vec{u}_1 + \vec{u}_2T)([F_{(1)}^2]F_{(1)} + [F_{(1)}\Ft_{(1)}]\Ft_{(1)}),
\ee
where we define
\be
\Ft_{(k)}^{\m\n}
\equiv\hlf\ove{\sqrt{-g}}\e^{\m\n\r\s}F_{(k)\r\s},\qquad
\Ft_{(k)\m\n} \equiv g_{\m\r}g_{\n\s}\Ft_{(k)}^{\r\s},
\ee
i.e. dualisation as well as raising and lowering of indices
are with respect to the metric $g_{\m\n}$ only.
Then from eq.\eq{tsd3_int2}, we obtain
\be\label{F3}
F_{(3)}
=
b_0e^{-2\f}(\bar{T}-T^*)^{-1}([F_{(1)}^2]\Ft_{(1)} - [F_{(1)}\Ft_{(1)}]F_{(1)}).
\ee
We may then follow the similar steps to analyse higher orders of eq.\eq{tsd_expand}.
Working with the fifth order term and using
\be
\begin{split}
\hat{F}_{(3)}
&=\bar{T}F_{(3)}\\
&=C_0 F_{(3)} + e^{-\f}\tilde{F}_{(3)} + b_0e^{-2\f}([F_{(1)}^2]\Ft_{(1)} - [F_{(1)}\Ft_{(1)}]F_{(1)}),
\end{split}
\ee
one obtains
\be\label{F5}
\begin{split}
F_{(5)}&=e^{-2\f} b_{0} (\bar{T}-T^*)^{-1}
\bigg(- e^{-\f} b_{0}[F_{(1)}^2] [F_{(1)}\tilde{F}_{(1)}] F_{(1)} 
 - 2 [F_{(1)}\tilde{F}_{(3)}] F_{(1)}\\
&\qquad
+\frac{3}{2} e^{-\f}b_{0} [F_{(1)}^2]^{2} \Ft_{(1)} +  \frac{1}{2} e^{-\f}b_0 [F_{(1)}\tilde{F}_{(1)}]^{2} \Ft_{(1)} + 2 [F_{(1)}F_{(3)}]\Ft_{(1)}\\
&\qquad+ [F_{(1)}^2]\Ft_{(3)}
- [F_{(1)}\tilde{F}_{(1)}] F_{(3)}
\bigg).
\end{split}
\ee
So in summary
\be\label{HXQX_nl}
\vec{H}
=(\vec{u}_1 + \vec{u}_2 T^*)F_{(1)}
+(\vec{u}_1 + \vec{u}_2 \bar{T})(F_{(3)} + F_{(5)})
+\cdots,
\ee
where $F_{(1)}, F_{(3)}, F_{(5)}$ are given by eq.\eq{F1}, \eq{F3}, \eq{F5}.
Next, by using $\vec{R}_X = \vec{Q}_X - \vec{H}$,
and $\vec{Q}_X = (\vec{u}_1 + \vec{u}_2\bar{T}^*)Q_X$ (due to $\vec{Q}_X = \bar{*}\vec{Q}_X$),
we obtain
\be\label{RXQX_nl}
\vec{R}_X
=(\vec{u}_1 + \bar{T}\vec{u}_2)
\lrbrk{(\bar{T} - T^*)^{-1}(\bar{T}^* - T^*)Q_X - F_{(3)} - F_{(5)}}
+\cdots.
\ee
Higher order expansions for $\vec{H}$ and $\vec{R}_X$
can also be obtained straightforwardly,
but it can be expected that
the higher the order, the more involved the expression become.
A better algorithm might be required to systematically extract higher orders.
This is left as a future work.

The analysis is much simpler
in the case of flat metrics.
In this case, the form of the expansion of $\vec{R}, \vec{H}$
in terms of independent fields
can be determined to any perturbative order.
We will set $\bar{g}_{\m\n} = g_{\m\n} = \h_{\m\n}$
and $C_0 = \f = \bar{C}_0 = \bar{\f} = 0$.
So $\bar{T} = T = \bar{\bullet} = \bullet$.
Furthermore, for simplicity, we will set $\vec{X} = \vec{0}$
which can easily be recovered later.
We have shown in subsection \ref{subsec:Rnl}
that the action \eq{nl-tSDSen} is invariant under $SL(2,\RR)$
duality transformation.
So after fixing the background fields,
the action is now invariant under
$SO(2)$ duality transformation
\be\label{SO2flat}
\vec{P}'=\vec{P}
\begin{pmatrix}
\cos\th & \sin\th\\
-\sin\th & \cos\th
\end{pmatrix}
,\qquad
\vec{Q}'=\vec{Q}
\begin{pmatrix}
\cos\th & \sin\th\\
-\sin\th & \cos\th
\end{pmatrix}
.
\ee
It is also clear that in the case of flat metrics,
the action \eq{nl-tSDSen} is invariant under standard Lorentz transformation.

Due to $\vec{Q} = (Q^1,Q^{\hat{1}}) = \bar{*}\vec{Q}$,
the relationship between 
$Q^1_{\m\n}$ and $Q^{\hat{1}}_{\m\n}$
is
$Q^{\hat{1}}_{\m\n} = \e_{\m\n\r\s}Q^{1\r\s}/2$.
There are only two independent Lorentz invariant quantities,
which are $Q^{1\m\n}Q^1_{\m\n}$ and $Q^{1\m\n}Q^{\hat{1}}_{\m\n}$.
These quantities however are not invariant under the duality transformation.
There is only one independent quantity which is both Lorentz invariant and duality symmetric.
This quantity is $(Q^{1\m\n}Q^1_{\m\n})^2 + (Q^{1\m\n}Q^{\hat{1}}_{\m\n})^2$.
Denote
\be\label{Q2}
(Q^2)^{IJ}\equiv Q^I_{\m\n}Q^{J\n\m},\quad
[Q^2]\equiv(Q^2)^{II},
\ee
\be\label{Q3}
(Q^3)^I_{\m\n}\equiv (Q^2)^{IJ} Q^J_{\m\n},\quad
\vec{Q}^3\equiv \hlf (Q^3)^I_{\n\m}\vec{u}_I dx^\m\w d x^\n,
\ee
\be\label{Q4}
(Q^4)^{IJ}\equiv (Q^2)^{IK}(Q^2)^{KJ},\quad
[Q^4]\equiv (Q^4)^{II}.
\ee
It can be seen that $[Q^4]\propto ((Q^{1\m\n}Q^1_{\m\n})^2 + (Q^{1\m\n}Q^{\hat{1}}_{\m\n})^2)$.
It is convenient to denote the invariant using
\be\label{Ydefn}
Y\equiv\frac{[Q^4]}{2}.
\ee
It can also be shown from $\vec{Q} = \bar{*}\vec{Q}$
that
the general form of quantities which are $2$-tuple $2$-forms is
\be
l_1(Y)\vec{Q} + l_2(Y)\vec{Q}^3.
\ee
Note that $\vec{Q}^3 = -\bar{*}\vec{Q}^3$.
So in general, $\vec{R}$ in the case of flat metrics takes the form
\be\label{RfQ}
\vec{R}
=l(Y)\vec{Q}^3.
\ee
By noting that $(Q^2)^{IJ}$ is a $2\times 2$ matrix
and that $[Q^2] = 0$, we obtain
Cayley-Hamilton formula for $(Q^2)^{IJ}$:
\be
(Q^4)^{IJ} = \hlf[Q^4]\d^{IJ}.
\ee
Then by using
$\vec{H} = \vec{Q} - \vec{R}$
and eq.\eq{RfQ},
we obtain
\be\label{starHflat}
*\vec{H}
=\frac{1+l^4 Y^2 + 6l^2 Y}{1-l^4Y^2}\vec{H} + \frac{2l}{1-l^4 Y^2}\vec{H}^3,
\ee
\be\label{trH2flat}
[H^2] = -4lY,
\ee
\be\label{trH4flat}
[H^4] = 2Y(1+l^4 Y^2 + 6l^2 Y),
\ee
where $(H^2)^{IJ}, \vec{H}^3, (H^4)^{IJ},$ etc.
are as defined in eq.\eq{H2_curved}-\eq{H3_curved}.
Note that the equation \eq{starHflat}
is already in the form of eq.\eq{twisted_sd_2}
with
\be\label{Kl_flat}
K = \frac{1-l^2 Y}{1+l^2 Y}.
\ee
Note from the analysis that only by using symmetry requirements,
one obtains the non-linear twisted self-duality conditions
which are automatically consistent with the constitutive relations.
On the other hand, we have seen previously in subsection \ref{subsec:TwistedSDSen} that
in the case of curved metrics, it is more involved
as the comparison with constitutive relations should explicitly be made.

As a summary so far, in the case of flat metric,
the composite fields $\vec{R}, \vec{H}$
can be expressed in terms of $\vec{Q}$ as
\be\label{RHQ_flat}
\vec{R}
=l(Y)\vec{Q}^3,\qquad
\vec{H}
=\vec{Q} - l(Y)\vec{Q}^3,
\ee
such that different $l$ corresponds to theories with different non-linear twisted self-duality condition \eq{starHflat}.
The values of $l$ which make eq.\eq{starHflat} singular
are $l=\pm\sqrt{1/Y}$ which correspond to the singular case $K = 0$,
c.f. the discussion below eq.\eq{twisted_sd_2}.

It can easily be shown directly that $\vec{R}$ and $\vec{H}$
as given in eq.\eq{RHQ_flat} transforms under $SO(2)$
duality rotation eq.\eq{SO2flat} as
\be\label{SO2flat_RH}
\vec{R}'=\vec{R}
\begin{pmatrix}
\cos\th & \sin\th\\
-\sin\th & \cos\th
\end{pmatrix}
,\qquad
\vec{H}'=\vec{H}
\begin{pmatrix}
\cos\th & \sin\th\\
-\sin\th & \cos\th
\end{pmatrix}
.
\ee

With flat metric, it is also possible
to express the non-linear duality-symmetric Sen action
entirely in terms of independent fields.
By applying $\Omega(\vec{Q}\w\d_Q\cdot)$ on eq.\eq{RfQ},
we obtain
\be\label{dQflat}
\d_Q
\lrbrk{\hlf\Omega\lrbrk{\vec{Q}\w\vec{R}}+
d^4 x\ U_0(Y)}
=\Omega(\d\vec{Q}\w\vec{R}),
\ee
where $U_0(Y)$ satisfies
\be\label{U0prime}
U_0'(Y) = \hlf Yl'(Y) + \ove{4}l(Y).
\ee
The action whose equations of motion imply $d\vec{H} = \vec{0}$
then reads
\be\label{Sflat_Q4}
S_{\textrm{Sen,flat}}
=-\int\Omega\lrbrk{\ove{4}d\vec{P}\w\bar{*}d\vec{P} - \vec{Q}\w d\vec{P} + \hlf\vec{Q}\w\vec{R}}
-\int d^4 x U_0(Y),
\ee
where $\vec{P}$
and $\vec{Q} = \bar{*}\vec{Q}$
are independent fields
and recall that $Y$ and $\vec{R}$ are given by eq.\eq{Ydefn} and eq.\eq{RfQ}.
By using eq.\eq{SO2flat} and eq.\eq{SO2flat_RH},
it can be shown that the action \eq{Sflat_Q4}
is invariant under $SO(2)$ duality transformation.
Alternatively, the action \eq{Sflat_Q4}
can be written as
\be\label{Sflat_Q4_alt}
S_{\textrm{Sen,flat}}
=-\int\Omega\lrbrk{\ove{4}d\vec{P}\w\bar{*}d\vec{P} - \vec{Q}\w d\vec{P}}
-\int d^4 x U_1(Y),
\ee
where $U_1(Y)$ satisfies
\be\label{U1prime}
U_1'(Y) = -\ove{4}l(Y).
\ee

As an example of a 
non-linear duality-symmetric Sen theory
with flat metrics,
consider the theory
corresponding to DBI theory.
In order to determine $l$ and $U_1$,
we use the expression of $K$ from eq.\eq{K_DBI}.
Then after using eq.\eq{trH2flat}, \eq{Kl_flat},
we arrive at
\be
l^4 Y^2 - 2 l^2 Y - 8l + 1 = 0,
\ee
whose regular series solution is given by
\be\label{l_DBI}
l
=\sum_{n=0}^\infty(-1)^n{4n+1 \choose n}\frac{1}{3n+2}\frac{1}{2^{6n + 2}}Y^n.
\ee
Then by using eq.\eq{U1prime}, we obtain
\be
U_1
=\sum_{n=0}^\infty(-1)^{n+1}{4n+1 \choose n}\frac{1}{3n+2}\frac{1}{2^{6n + 4}}\frac{Y^{n+1}}{n+1},
\ee
which can be substituted into eq.\eq{Sflat_Q4_alt}
to obtain
a non-linear duality-symmetric Sen action
with flat metrics corresponding to DBI theory.

\section{Conclusion and Discussion}\label{sec:conclusion}
In this paper, we constructed the Sen action \eq{nl-tSDSen} for non-linear $SL(2,\RR)$ duality-symmetric theories and
an $SL(2,\RR)$ duality-symmetric D3-brane action \eq{SenD3} in the Sen formalism coupled to type IIB supergravity background.
The constructed actions lead to expected equations of motion
for the systems with non-linear $SL(2,\RR)$ duality symmetry
and the D3-brane coupled to type IIB supergravity background.
$SL(2,\RR)$ duality-symmetry and diffeomorphism of the actions
are shown explicitly. Furthermore, the kappa symmetry of the D3-brane action
is also shown.

For each different non-linear theory with action \eq{nl-tSDSen}, the function $U$ appearing in the action is different. Furthermore, the composite 2-tuple 2-form field $\vec{H}$ satisfies off-shell non-linear twisted self-duality condition \eq{twisted_sd_2} which is also different for different theories.
We have explained in subsection \ref{subsec:TwistedSDSen}
that the condition \eq{twisted_sd_2}
arises from the requirement that
non-linear twisted-self-duality condition
should be equivalent to the constitutive relation \eq{consti_gen}.

The composite field $\vec{H}$ is exact on-shell in the spacetime with trivial topology. Therefore, $\vec{H}$ corresponds to the degrees of freedom of gauge field doublet whose equation of motion is given by non-linear twisted-self-duality condition. 

The $SL(2,\RR)$ duality-symmetry of the action \eq{nl-tSDSen}
is proved for the quadratic case in subsection \ref{subsec:quad_TwistedSDSen}
and perturbatively for the non-linear cases in subsection \ref{subsec:Rnl}.
In both of these cases,
while the independent fields $\vec{P}, \vec{Q}$
transform under the standard $SL(2,\RR)$ duality transformation,
it is a priori unclear how the composite fields $\vec{R}$ and $\vec{H}$ transform.
We have shown that
the fields $\vec{R}$ and $\vec{H}$ also transform under the $SL(2,\RR)$ standard duality transformation.
It is then easy to see that the actions are $SL(2,\RR)$ duality invariant.

As part of the features of the Sen formalism,
the system described by the action \eq{nl-tSDSen}
is in fact separated into two sectors, uncoupled from each other.
The physical $g$-sector describes the gauge field doublet
whose field strength is non-linearly twisted-self-dual (according to eq.\eq{twisted_sd_os}). It is coupled to the external fields
which are the metric $g_{\m\n}$,
an axion-dilaton pair $\t$, and the $2$-tuple $2$-form field $\vec{X}$.
The other sector is the unphysical $\bar{g}$-sector
which describes another gauge field doublet
whose field strength is linearly twisted-self-dual (according to eq.\eq{xiplusP}).
It is coupled to the external fields
which are the other metric $\bar{g}_{\m\n}$,
and the other axion-dilaton pair $\bar{\t}$.
The separation into two sectors is verified
both at the equations of motion level
as discussed in subsections \ref{subsec:quad_TwistedSDSen}-\ref{subsec:TwistedSDSen}
and at the Hamiltonian level
in subsection \ref{subsec:Ham}.
The Hamiltonian analysis
also justifies that the physical $g$-sector
has the correct sign of the kinetic energy
whereas the unphysical $\bar{g}$-sector has the incorrect sign of the kinetic energy.

The action \eq{nl-tSDSen} also has three related types of diffeomorphisms.
As shown in subsection \ref{subsec:diffeo},
the most non-trivial check is the $\zeta$-symmetry
where the fields in the physical sector transform under the standard diffeomorphism
whereas fields in the unphysical sector do not transform.

The form of
non-linear twisted self-duality condition and $U$ can be solved perturbatively
starting from the expansion \eq{Kpert}.
This gives rise to eq.\eq{Vpert}-\eq{Upert}.
The corresponding non-linear Sen action describes
the system whose dynamics of the physical sector
can also be described by the one-gauge-field Lagrangian \eq{one_field_pert}.

We have shown that an appropriate form of $U$
can be chosen in the action \eq{nl-tSDSen}
so that it can be extended to be an $SL(2,\RR)$ duality-symmetric action \eq{SenD3} for D3-brane coupled to type IIB supergravity.
The embedding is in the Green-Schwarz formalism
where the bosonic worldvolume is embedded in the background target superspace.
An important verification for any action in the Green-Schwarz formalism
is to show that it has kappa-symmetry which can be used to realise the matching of the number of bosonic and fermionic degrees of freedom.
We have shown in subsection \ref{subsec:SenD3}
that the action \eq{SenD3} indeed has kappa-symmetry.

A natural extension to the works in this paper would be to construct duality-symmetric actions in the Sen formalism
describing the system whose physical sector
has $n>1$ pairs of physical $2$-form fields
satisfying twisted self-duality condition.
As concrete examples, one might construct actions in the Sen formalism
giving equations of motion agreeing with \cite{Cremmer:1979up}, \cite{deWit:1982bul}, \cite{Castellani:1985ka}, \cite{Kallosh:2008ic}.

Although this paper only discusses the case $n = 1$, the form of the equations hint that, at least for the linear twisted self-duality case, the extensions to $n>1$ is quite straightforward, e.g. simply by extending the index $I$ to take values in $\{1,2,\cdots,n,\hat{1},\hat{2},\cdots,\hat{n}\}$.
The task to follow would be to complete the supersymmetry or supergravity extensions to obtain the required $n>1$ duality-symmetric theories in the Sen formalism.

\section*{Acknowledgments}
We would like to thank Martin Cederwall and Sheng-Lan Ko
for discussions.
This research has received funding support from the NSRF via the Program Management Unit for Human Resources \& Institutional Development, Research and Innovation [grant number B39G680009].

\providecommand{\href}[2]{#2}\begingroup\raggedright\endgroup


\begin{thebibliography}{10}

\bibitem{Gaillard:1981rj}
M.K.~Gaillard and B.~Zumino, \emph{{Duality Rotations for Interacting Fields}},
  \href{https://doi.org/10.1016/0550-3213(81)90527-7}{\emph{Nucl. Phys. B}
  {\bfseries 193} (1981) 221}.

\bibitem{Gibbons:1995cv}
G.W.~Gibbons and D.A.~Rasheed, \emph{{Electric - magnetic duality rotations in
  nonlinear electrodynamics}},
  \href{https://doi.org/10.1016/0550-3213(95)00409-L}{\emph{Nucl. Phys. B}
  {\bfseries 454} (1995) 185}
  [\href{https://arxiv.org/abs/hep-th/9506035}{{\ttfamily hep-th/9506035}}].

\bibitem{Ceresole:1995jg}
A.~Ceresole, R.~D'Auria, S.~Ferrara and A.~Van~Proeyen, \emph{{Duality
  transformations in supersymmetric Yang-Mills theories coupled to
  supergravity}},
  \href{https://doi.org/10.1016/0550-3213(95)00175-R}{\emph{Nucl. Phys. B}
  {\bfseries 444} (1995) 92}
  [\href{https://arxiv.org/abs/hep-th/9502072}{{\ttfamily hep-th/9502072}}].

\bibitem{Gibbons:1995ap}
G.W.~Gibbons and D.A.~Rasheed, \emph{{Sl(2,R) invariance of nonlinear
  electrodynamics coupled to an axion and a dilaton}},
  \href{https://doi.org/10.1016/0370-2693(95)01272-9}{\emph{Phys. Lett. B}
  {\bfseries 365} (1996) 46}
  [\href{https://arxiv.org/abs/hep-th/9509141}{{\ttfamily hep-th/9509141}}].

\bibitem{Gaillard:1997rt}
M.K.~Gaillard and B.~Zumino, \emph{{Nonlinear electromagnetic selfduality and
  Legendre transformations}},  in \emph{{A Newton Institute Euroconference on
  Duality and Supersymmetric Theories}}, pp.~33--48, 12, 1997
  [\href{https://arxiv.org/abs/hep-th/9712103}{{\ttfamily hep-th/9712103}}].

\bibitem{Andrianopoli:1996bq}
L.~Andrianopoli, R.~D'Auria, S.~Ferrara, P.~Fre and M.~Trigiante, \emph{{RR
  scalars, U duality and solvable Lie algebras}},
  \href{https://doi.org/10.1016/S0550-3213(97)00220-4}{\emph{Nucl. Phys. B}
  {\bfseries 496} (1997) 617}
  [\href{https://arxiv.org/abs/hep-th/9611014}{{\ttfamily hep-th/9611014}}].

\bibitem{Andrianopoli:1996ve}
L.~Andrianopoli, R.~D'Auria and S.~Ferrara, \emph{{U duality and central
  charges in various dimensions revisited}},
  \href{https://doi.org/10.1142/S0217751X98000196}{\emph{Int. J. Mod. Phys. A}
  {\bfseries 13} (1998) 431}
  [\href{https://arxiv.org/abs/hep-th/9612105}{{\ttfamily hep-th/9612105}}].

\bibitem{Gaillard:1997zr}
M.K.~Gaillard and B.~Zumino, \emph{{Selfduality in nonlinear
  electromagnetism}}, \href{https://doi.org/10.1007/BFb0105236}{\emph{Lect.
  Notes Phys.} {\bfseries 509} (1998) 121}
  [\href{https://arxiv.org/abs/hep-th/9705226}{{\ttfamily hep-th/9705226}}].

\bibitem{Araki:1998nn}
M.~Araki and Y.~Tanii, \emph{{Duality symmetries in nonlinear gauge theories}},
  \href{https://doi.org/10.1142/S0217751X99000579}{\emph{Int. J. Mod. Phys. A}
  {\bfseries 14} (1999) 1139}
  [\href{https://arxiv.org/abs/hep-th/9808029}{{\ttfamily hep-th/9808029}}].

\bibitem{Aschieri:2008ns}
P.~Aschieri, S.~Ferrara and B.~Zumino, \emph{{Duality Rotations in Nonlinear
  Electrodynamics and in Extended Supergravity}},
  \href{https://doi.org/10.1393/ncr/i2008-10038-8}{\emph{Riv. Nuovo Cim.}
  {\bfseries 31} (2008) 625} [\href{https://arxiv.org/abs/0807.4039}{{\ttfamily
  0807.4039}}].

\bibitem{Zwanziger:1970hk}
D.~Zwanziger, \emph{{Local Lagrangian quantum field theory of electric and
  magnetic charges}}, \href{https://doi.org/10.1103/PhysRevD.3.880}{\emph{Phys.
  Rev. D} {\bfseries 3} (1971) 880}.

\bibitem{Deser:1976iy}
S.~Deser and C.~Teitelboim, \emph{{Duality Transformations of Abelian and
  Nonabelian Gauge Fields}},
  \href{https://doi.org/10.1103/PhysRevD.13.1592}{\emph{Phys. Rev. D}
  {\bfseries 13} (1976) 1592}.

\bibitem{Pasti:1995ii}
P.~Pasti, D.P.~Sorokin and M.~Tonin, \emph{{Note on manifest Lorentz and
  general coordinate invariance in duality symmetric models}},
  \href{https://doi.org/10.1016/0370-2693(95)00463-U}{\emph{Phys. Lett. B}
  {\bfseries 352} (1995) 59}
  [\href{https://arxiv.org/abs/hep-th/9503182}{{\ttfamily hep-th/9503182}}].

\bibitem{Pasti:1995tn}
P.~Pasti, D.P.~Sorokin and M.~Tonin, \emph{{Duality symmetric actions with
  manifest space-time symmetries}},
  \href{https://doi.org/10.1103/PhysRevD.52.R4277}{\emph{Phys. Rev. D}
  {\bfseries 52} (1995) R4277}
  [\href{https://arxiv.org/abs/hep-th/9506109}{{\ttfamily hep-th/9506109}}].

\bibitem{Maznytsia:1998xw}
A.~Maznytsia, C.R.~Preitschopf and D.P.~Sorokin, \emph{{Duality of selfdual
  actions}}, \href{https://doi.org/10.1016/S0550-3213(98)00741-X}{\emph{Nucl.
  Phys. B} {\bfseries 539} (1999) 438}
  [\href{https://arxiv.org/abs/hep-th/9805110}{{\ttfamily hep-th/9805110}}].

\bibitem{Bekaert:2001wa}
X.~Bekaert and S.~Cucu, \emph{{Deformations of duality symmetric theories}},
  \href{https://doi.org/10.1016/S0550-3213(01)00260-7}{\emph{Nucl. Phys. B}
  {\bfseries 610} (2001) 433}
  [\href{https://arxiv.org/abs/hep-th/0104048}{{\ttfamily hep-th/0104048}}].

\bibitem{Bossard:2011ij}
G.~Bossard and H.~Nicolai, \emph{{Counterterms vs. Dualities}},
  \href{https://doi.org/10.1007/JHEP08(2011)074}{\emph{JHEP} {\bfseries 08}
  (2011) 074} [\href{https://arxiv.org/abs/1105.1273}{{\ttfamily 1105.1273}}].

\bibitem{Pasti:2012wv}
P.~Pasti, D.~Sorokin and M.~Tonin, \emph{{Covariant actions for models with
  non-linear twisted self-duality}},
  \href{https://doi.org/10.1103/PhysRevD.86.045013}{\emph{Phys. Rev. D}
  {\bfseries 86} (2012) 045013}
  [\href{https://arxiv.org/abs/1205.4243}{{\ttfamily 1205.4243}}].

\bibitem{Lee:2013ewa}
C.~Lee and H.~Min, \emph{{SL(2,R) duality-symmetric action for electromagnetic
  theory with electric and magnetic sources}},
  \href{https://doi.org/10.1016/j.aop.2013.09.015}{\emph{Annals Phys.}
  {\bfseries 339} (2013) 328}
  [\href{https://arxiv.org/abs/1306.5520}{{\ttfamily 1306.5520}}].

\bibitem{Cederwall:1997ab}
M.~Cederwall and A.~Westerberg, \emph{{World volume fields, SL(2:Z) and
  duality: The Type IIB three-brane}},
  \href{https://doi.org/10.1088/1126-6708/1998/02/004}{\emph{JHEP} {\bfseries
  02} (1998) 004} [\href{https://arxiv.org/abs/hep-th/9710007}{{\ttfamily
  hep-th/9710007}}].

\bibitem{Cederwall:1997ts}
M.~Cederwall and P.K.~Townsend, \emph{{The Manifestly Sl(2,Z) covariant
  superstring}},
  \href{https://doi.org/10.1088/1126-6708/1997/09/003}{\emph{JHEP} {\bfseries
  09} (1997) 003} [\href{https://arxiv.org/abs/hep-th/9709002}{{\ttfamily
  hep-th/9709002}}].

\bibitem{Berman:1997iz}
D.~Berman, \emph{{SL(2,Z) duality of Born-Infeld theory from nonlinear selfdual
  electrodynamics in six-dimensions}},
  \href{https://doi.org/10.1016/S0370-2693(97)00919-2}{\emph{Phys. Lett. B}
  {\bfseries 409} (1997) 153}
  [\href{https://arxiv.org/abs/hep-th/9706208}{{\ttfamily hep-th/9706208}}].

\bibitem{Nurmagambetov:1998gp}
A.~Nurmagambetov, \emph{{Duality symmetric three-brane and its coupling to type
  IIB supergravity}},
  \href{https://doi.org/10.1016/S0370-2693(98)00848-X}{\emph{Phys. Lett. B}
  {\bfseries 436} (1998) 289}
  [\href{https://arxiv.org/abs/hep-th/9804157}{{\ttfamily hep-th/9804157}}].

\bibitem{Suzuki:1999aa}
T.~Suzuki, \emph{{Supersymmetric action of SL(2:Z) covariant D3-brane and its
  kappa symmetry}},
  \href{https://doi.org/10.1016/S0370-2693(00)00114-3}{\emph{Phys. Lett. B}
  {\bfseries 476} (2000) 387}
  [\href{https://arxiv.org/abs/hep-th/9911201}{{\ttfamily hep-th/9911201}}].

\bibitem{Vanichchapongjaroen:2017zuh}
P.~Vanichchapongjaroen, \emph{{Dual formulation of covariant nonlinear
  duality-symmetric action of kappa-symmetric D3-brane}},
  \href{https://doi.org/10.1007/JHEP02(2018)116}{\emph{JHEP} {\bfseries 02}
  (2018) 116} [\href{https://arxiv.org/abs/1712.06425}{{\ttfamily
  1712.06425}}].

\bibitem{Avetisyan:2021heg}
Z.~Avetisyan, O.~Evnin and K.~Mkrtchyan, \emph{{Democratic Lagrangians for
  Nonlinear Electrodynamics}},
  \href{https://doi.org/10.1103/PhysRevLett.127.271601}{\emph{Phys. Rev. Lett.}
  {\bfseries 127} (2021) 271601}
  [\href{https://arxiv.org/abs/2108.01103}{{\ttfamily 2108.01103}}].

\bibitem{Mkrtchyan:2022xrm}
K.~Mkrtchyan and F.~Valach, \emph{{Democratic actions for type II
  supergravities}},
  \href{https://doi.org/10.1103/PhysRevD.107.066027}{\emph{Phys. Rev. D}
  {\bfseries 107} (2023) 066027}
  [\href{https://arxiv.org/abs/2207.00626}{{\ttfamily 2207.00626}}].

\bibitem{Sen:2015nph}
A.~Sen, \emph{{Covariant Action for Type IIB Supergravity}},
  \href{https://doi.org/10.1007/JHEP07(2016)017}{\emph{JHEP} {\bfseries 07}
  (2016) 017} [\href{https://arxiv.org/abs/1511.08220}{{\ttfamily
  1511.08220}}].

\bibitem{Sen:2019qit}
A.~Sen, \emph{{Self-dual forms: Action, Hamiltonian and Compactification}},
  \href{https://doi.org/10.1088/1751-8121/ab5423}{\emph{J. Phys. A} {\bfseries
  53} (2020) 084002} [\href{https://arxiv.org/abs/1903.12196}{{\ttfamily
  1903.12196}}].

\bibitem{Sen:2015uaa}
A.~Sen, \emph{{BV Master Action for Heterotic and Type II String Field
  Theories}}, \href{https://doi.org/10.1007/JHEP02(2016)087}{\emph{JHEP}
  {\bfseries 02} (2016) 087}
  [\href{https://arxiv.org/abs/1508.05387}{{\ttfamily 1508.05387}}].

\bibitem{Lambert:2019diy}
N.~Lambert, \emph{{(2,0) Lagrangian Structures}},
  \href{https://doi.org/10.1016/j.physletb.2019.134948}{\emph{Phys. Lett. B}
  {\bfseries 798} (2019) 134948}
  [\href{https://arxiv.org/abs/1908.10752}{{\ttfamily 1908.10752}}].

\bibitem{Andriolo:2020ykk}
E.~Andriolo, N.~Lambert and C.~Papageorgakis, \emph{{Geometrical Aspects of An
  Abelian (2,0) Action}},
  \href{https://doi.org/10.1007/JHEP04(2020)200}{\emph{JHEP} {\bfseries 04}
  (2020) 200} [\href{https://arxiv.org/abs/2003.10567}{{\ttfamily
  2003.10567}}].

\bibitem{Vanichchapongjaroen:2020wza}
P.~Vanichchapongjaroen, \emph{{Covariant M5-brane action with self-dual
  3-form}}, \href{https://doi.org/10.1007/JHEP05(2021)039}{\emph{JHEP}
  {\bfseries 05} (2021) 039}
  [\href{https://arxiv.org/abs/2011.14384}{{\ttfamily 2011.14384}}].

\bibitem{Evnin:2022kqn}
O.~Evnin and K.~Mkrtchyan, \emph{{Three approaches to chiral form
  interactions}},
  \href{https://doi.org/10.1016/j.difgeo.2023.102016}{\emph{Differ. Geom.
  Appl.} {\bfseries 89} (2023) 102016}
  [\href{https://arxiv.org/abs/2207.01767}{{\ttfamily 2207.01767}}].

\bibitem{Andrianopoli:2022bzr}
L.~Andrianopoli, C.A.~Cremonini, R.~D'Auria, P.A.~Grassi, R.~Matrecano,
  R.~Noris et~al., \emph{{M5-brane in the superspace approach}},
  \href{https://doi.org/10.1103/PhysRevD.106.026010}{\emph{Phys. Rev. D}
  {\bfseries 106} (2022) 026010}
  [\href{https://arxiv.org/abs/2206.06388}{{\ttfamily 2206.06388}}].

\bibitem{Lambert:2023qgs}
N.~Lambert, \emph{{Duality and fluxes in the sen formulation of self-dual
  fields}}, \href{https://doi.org/10.1016/j.physletb.2023.137888}{\emph{Phys.
  Lett. B} {\bfseries 840} (2023) 137888}
  [\href{https://arxiv.org/abs/2302.10955}{{\ttfamily 2302.10955}}].

\bibitem{Phonchantuek:2023iao}
A.~Phonchantuek and P.~Vanichchapongjaroen, \emph{{Double dimensional reduction
  of M5-brane action in Sen formalism}},
  \href{https://doi.org/10.1140/epjc/s10052-023-11892-2}{\emph{Eur. Phys. J. C}
  {\bfseries 83} (2023) 721}
  [\href{https://arxiv.org/abs/2305.04861}{{\ttfamily 2305.04861}}].

\bibitem{Chakrabarti:2023czz}
S.~Chakrabarti and M.~Raman, \emph{{Exploring T-Duality for Self-Dual Fields}},
  \href{https://doi.org/10.1002/prop.202400023}{\emph{Fortsch. Phys.}
  {\bfseries 72} (2024) 2400023}
  [\href{https://arxiv.org/abs/2311.09153}{{\ttfamily 2311.09153}}].

\bibitem{Hull:2023dgp}
C.M.~Hull, \emph{{Covariant action for self-dual p-form gauge fields in general
  spacetimes}}, \href{https://doi.org/10.1007/JHEP04(2024)011}{\emph{JHEP}
  {\bfseries 04} (2024) 011}
  [\href{https://arxiv.org/abs/2307.04748}{{\ttfamily 2307.04748}}].

\bibitem{Aggarwal:2025fiq}
A.~Aggarwal, S.~Chakrabarti and M.~Raman, \emph{{Monopoles, Clarified}},
  \href{https://arxiv.org/abs/2504.16673}{{\ttfamily 2504.16673}}.

\bibitem{Cederwall:1996ri}
M.~Cederwall, A.~von Gussich, B.E.W.~Nilsson, P.~Sundell and A.~Westerberg,
  \emph{{The Dirichlet super p-branes in ten-dimensional type IIA and IIB
  supergravity}},
  \href{https://doi.org/10.1016/S0550-3213(97)00075-8}{\emph{Nucl. Phys. B}
  {\bfseries 490} (1997) 179}
  [\href{https://arxiv.org/abs/hep-th/9611159}{{\ttfamily hep-th/9611159}}].

\bibitem{Cederwall:1996pv}
M.~Cederwall, A.~von Gussich, B.E.W.~Nilsson and A.~Westerberg, \emph{{The
  Dirichlet super three-brane in ten-dimensional type IIB supergravity}},
  \href{https://doi.org/10.1016/S0550-3213(97)00071-0}{\emph{Nucl. Phys. B}
  {\bfseries 490} (1997) 163}
  [\href{https://arxiv.org/abs/hep-th/9610148}{{\ttfamily hep-th/9610148}}].

\bibitem{Janaun:2024wya}
S.~Janaun, A.~Phonchantuek and P.~Vanichchapongjaroen, \emph{{Nonlinear chiral
  forms in the Sen formulation}},
  \href{https://doi.org/10.1140/epjc/s10052-024-13207-5}{\emph{Eur. Phys. J. C}
  {\bfseries 84} (2024) 832}
  [\href{https://arxiv.org/abs/2404.05380}{{\ttfamily 2404.05380}}].

\bibitem{Cremmer:1979up}
E.~Cremmer and B.~Julia, \emph{{The SO(8) Supergravity}},
  \href{https://doi.org/10.1016/0550-3213(79)90331-6}{\emph{Nucl. Phys. B}
  {\bfseries 159} (1979) 141}.

\bibitem{deWit:1982bul}
B.~de~Wit and H.~Nicolai, \emph{{N=8 Supergravity}},
  \href{https://doi.org/10.1016/0550-3213(82)90120-1}{\emph{Nucl. Phys. B}
  {\bfseries 208} (1982) 323}.

\bibitem{Castellani:1985ka}
L.~Castellani, A.~Ceresole, S.~Ferrara, R.~D'Auria, P.~Fre and E.~Maina,
  \emph{{The Complete $N=3$ Matter Coupled Supergravity}},
  \href{https://doi.org/10.1016/0550-3213(86)90157-4}{\emph{Nucl. Phys. B}
  {\bfseries 268} (1986) 317}.

\bibitem{Kallosh:2008ic}
R.~Kallosh and M.~Soroush, \emph{{Explicit Action of E(7)(7) on N=8
  Supergravity Fields}},
  \href{https://doi.org/10.1016/j.nuclphysb.2008.04.006}{\emph{Nucl. Phys. B}
  {\bfseries 801} (2008) 25} [\href{https://arxiv.org/abs/0802.4106}{{\ttfamily
  0802.4106}}].

\end{thebibliography}
\end{document}